%% file: main.tex
\documentclass[sigconf, nonacm]{acmart}

\newcommand\vldbdoi{10.14778/3773749.3773758}
\newcommand\vldbpages{196 - 209}
\newcommand\vldbvolume{19}
\newcommand\vldbissue{2}
\newcommand\vldbyear{2025}
\newcommand\vldbauthors{\authors}
\newcommand\vldbtitle{\shorttitle} 
\newcommand\vldbavailabilityurl{https://github.com/DataManagementLab/llmeval-enterprise-challenges}
\newcommand\vldbpagestyle{empty} 

\usepackage{listings}
\usepackage{xfrac}
\usepackage{tabularx}
\usepackage{caption}
\usepackage{subcaption}
\usepackage{csquotes}
\usepackage{bm}
\usepackage[para]{threeparttable}
\usepackage{enumitem}
\usepackage{hyperref}
\usepackage{multirow}
\usepackage{array}
\usepackage{makecell}
\usepackage{siunitx}
\usepackage{graphicx}
\usepackage{pifont}

\begin{document}
\newcommand{\revised}[1]{\textcolor{blue}{#1}}
\newcommand{\cmark}{\textcolor{green!80!black}{\ding{51}}}

\title{Unveiling Challenges for LLMs in Enterprise Data Engineering}
\subtitle{With Extended Technical Report}

\author{Jan-Micha Bodensohn}
\orcid{0000-0003-4884-0300}
\affiliation{
  \institution{DFKI \& Technical University of Darmstadt}
}
\authornote{Authors with equal contribution, alphabetical ordering.
Correspondence goes to \href{mailto:jan-micha.bodensohn@cs.tu-darmstadt.de}{jan-micha.bodensohn@cs.tu-darmstadt.de}}

\author{Ulf Brackmann}
\orcid{0009-0000-6766-1362}
\affiliation{
  \institution{SAP SE \& DFKI}
}
\authornotemark[1]

\author{Liane Vogel}
\orcid{0000-0001-9768-8873}
\affiliation{
  \institution{Technical University of Darmstadt}
}
\authornotemark[1]

\author{Anupam Sanghi}
\orcid{0000-0003-4764-3583}
\affiliation{
  \institution{Indian Institute of Technology Hyderabad}
}
\authornote{Work done while the author was at Technical University of Darmstadt.}

\author{Carsten Binnig}
\orcid{0000-0002-2744-7836}
\affiliation{
  \institution{Technical University of Darmstadt\\\& DFKI}
}

\begin{abstract}
Large Language Models (LLMs) promise to automate data engineering on tabular data, offering enterprises a valuable opportunity to cut the high costs of manual data handling.
But the enterprise domain comes with unique challenges that existing LLM-based approaches for data engineering often overlook, such as large table sizes, more complex tasks, and the need for internal knowledge.
To bridge these gaps, we identify key enterprise-specific challenges related to data, tasks, and background knowledge and extensively evaluate how they affect data engineering with LLMs.
Our analysis reveals that LLMs face substantial limitations in real-world enterprise scenarios, with accuracy declining sharply.
Our findings contribute to a systematic understanding of LLMs for enterprise data engineering to support their adoption in industry.
\end{abstract}

\maketitle

\pagestyle{\vldbpagestyle}
\begingroup\small\noindent\raggedright\textbf{PVLDB Reference Format:}\\
\vldbauthors. \vldbtitle. PVLDB, \vldbvolume(\vldbissue): \vldbpages, \vldbyear.\\
\href{https://doi.org/\vldbdoi}{doi:\vldbdoi}
\endgroup
\begingroup
\renewcommand\thefootnote{}\footnote{\noindent
This work is licensed under the Creative Commons BY-NC-ND 4.0 International License. Visit \url{https://creativecommons.org/licenses/by-nc-nd/4.0/} to view a copy of this license. For any use beyond those covered by this license, obtain permission by emailing \href{mailto:info@vldb.org}{info@vldb.org}. Copyright is held by the owner/author(s). Publication rights licensed to the VLDB Endowment. \\
\raggedright Proceedings of the VLDB Endowment, Vol. \vldbvolume, No. \vldbissue\ %
ISSN 2150-8097. \\
\href{https://doi.org/\vldbdoi}{doi:\vldbdoi} \\
}\addtocounter{footnote}{-1}\endgroup

\ifdefempty{\vldbavailabilityurl}{}{
\vspace{.3cm}
\begingroup\small\noindent\raggedright\textbf{PVLDB Artifact Availability:}\\
The source code, data, and/or other artifacts have been made available at \url{\vldbavailabilityurl}.
\endgroup
}

\definecolor{COLOR_BLACK}{HTML}{000000}
\definecolor{COLOR_WHITE}{HTML}{FFFFFF}
\definecolor{COLOR_1A}{HTML}{5D85C3}
\definecolor{COLOR_2A}{HTML}{009CDA}
\definecolor{COLOR_3A}{HTML}{50B695}
\definecolor{COLOR_4A}{HTML}{AFCC50}
\definecolor{COLOR_5A}{HTML}{DDDF48}
\definecolor{COLOR_6A}{HTML}{FFE05C}
\definecolor{COLOR_7A}{HTML}{F8BA3C}
\definecolor{COLOR_8A}{HTML}{EE7A34}
\definecolor{COLOR_9A}{HTML}{E9503E}
\definecolor{COLOR_10A}{HTML}{C9308E}
\definecolor{COLOR_11A}{HTML}{804597}

\hyphenation{Sports-Tables}
\hyphenation{data-bases}
\hyphenation{data-base}
\newcommand{\sapcta}{\mbox{SAP\scriptsize{CTA}} }
\newcommand{\notesize}{\fontsize{7.5}{9}\selectfont}
\renewcommand\theadfont{\bfseries\notesize}

\renewcommand{\paragraph}[1]{\vspace{0.8ex}\noindent\textbf{#1}$\:$}
\newcommand{\newparagraph}{\vspace{0.8ex}\noindent}
\newcommand{\newparagraphwithindent}{\vspace{0.8ex}}
\newcommand{\challenge}[1]{\vspace{0.8ex}\noindent\textbf{#1}$\:$}
\newcommand{\experiment}[1]{\vspace{0.8ex}\noindent\textbf{#1}$\:$}

\input{sections_01_introduction}
\input{sections_02_background}
\input{sections_03_method}
\input{sections_04_data}
\input{sections_05_tasks}
\input{sections_06_knowledge}
\input{sections_07_costs}
\input{sections_08_conclusion}

\begin{acks}
This work has been supported by the BMBF and the state of Hesse as part of the NHR program and the HMWK cluster project 3AI.
It was also partially funded by the LOEWE Spitzenprofessur of the state of Hesse and has benefited from early stages of funding from the Deutsche Forschungsgemeinschaft (DFG) under Germany’s Excellence Strategy (EXC-3057); funding will begin in 2026.
 We thank DFKI Darmstadt, SAP, and hessian.AI as well as Alexey Streltsov, Matthias Urban, and Furkan Karakocaoglu for their support.
\end{acks}

\balance{}

\newpage

\bibliographystyle{ACM-Reference-Format}
\bibliography{bibliography}

\newpage
\nobalance
\input{appendix}

\end{document}

%% file: sections_01_introduction.tex
\section{Introduction}
\label{sec:introduction}

Large enterprises generate vast amounts of tabular data that drives applications like machine learning and analytical query processing.
Data engineering is crucial for understanding this raw data and preparing it for its downstream usage.
It encompasses a range of tasks, from data exploration and integration to data transformation and cleaning.
Since these tasks often impose significant manual overhead to apply existing tools to the specific data at hand, the automation of individual data engineering tasks like entity matching~\cite{konda_magellan_2016,mudgal2018deep} and column type annotation~\cite{hulsebos_sherlock_2019,zhang_sato_2020} with the help of machine learning has long drawn attention from researchers.
Nevertheless, adapting such machine learning approaches to new datasets and tasks often requires computer science expertise, rendering them inaccessible to many practitioners.

\newparagraphwithindent
Recent work has shown that Large Language Models~(LLMs) can be directly applied to data engineering tasks on tabular data, indicating that they achieve state-of-the-art results on various table-based tasks without requiring task-specific architectures and training~\cite{narayan_can_2022, jaimovitch-lopez_can_2022, kayali_chorus_2024, barke_solving_2024}.
Their out-of-the-box nature provides a significant advantage over other machine learning approaches that require supervised training for each dataset and task.
One example is the task of column type annotation, where the goal is to annotate the columns of a relational table with semantic types from a given ontology.
Whereas machine learning-based approaches like Sherlock~\cite{hulsebos_sherlock_2019} and Sato~\cite{zhang_sato_2020} require re-training for each new set of semantic types, LLMs can easily support different sets of semantic types by including them in the prompt~\cite{korini_column_2023}.
Therefore, they provide a promising avenue to automate data engineering tasks.

\begin{figure}
  \centering
  \includegraphics[width=\linewidth]{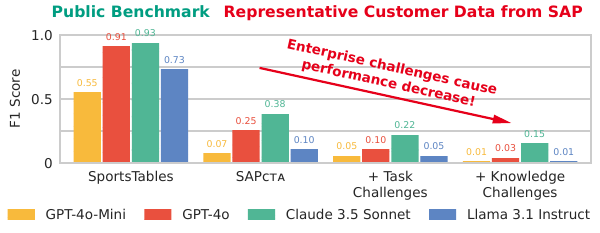}
    \vspace{-1.5em}
  \caption{LLMs perform well on public benchmarks but poorly in real-world enterprise settings.
  The plot compares support-weighted F1 scores for column type annotation on the public SportsTables benchmark with customer data from SAP.
  We further raise the difficulty \mbox{\emph{(+ Task Challenges)}} by increasing the number of semantic types from 200 (comparable to public benchmarks) to the full 5,089 from the enterprise setting and observe an additional performance drop.
   When also requiring internal knowledge about company-specific schema extensions in the form of customer-defined columns \mbox{\emph{(+ Knowledge Challenges)}}, the performance is close to zero.
  }
  \label{fig:headline_exp}
  \vspace{-1.2em}
\end{figure}

\paragraph{Enterprise challenges.}
Some first papers have now made the point that while LLMs achieve strong results on existing benchmarks, they fall short when applied to real-world enterprise data~\cite{kayali_mind_2025, bodensohn_llms_2024}.
A central observation here is that the data in public benchmarks often comes from web resources like Wikipedia~\cite{bhagavatula_tabel_2015} and GitHub~\cite{hulsebos_gittables_2023}.
In contrast, the data from companies running their business processes with software systems like those from SAP differs fundamentally from these datasets in many aspects, including table sizes, sparsity, and data types~\cite{kayali_mind_2025,savelieva_need_2022, vogelsgesang_get_2018}.
Since LLMs are typically trained on public data scraped from the web \cite{brown_language_2020,meta_llama_2024}, they have not seen much of this enterprise data during their training.

\newparagraphwithindent
In this paper, we set out to systematically study the performance of LLMs for enterprise data engineering on representative customer data and case studies reflecting real enterprise scenarios.
As a sneak peak into our results, Figure~\ref{fig:headline_exp} presents a first experiment comparing the performance of LLMs for column type annotation on a public benchmark to a dataset of real-world customer data from SAP.
As shown by the left bar group, LLMs of different types and sizes achieve high F1 scores when evaluated on a public benchmark like the SportsTables dataset~\cite{langenecker_sportstables_2023}.
The same models, however, show substantial performance decreases when applied to the representative customer data from SAP for column type annotation with a comparable number of semantic types, as shown in Figure~\ref{fig:headline_exp}~\mbox{\emph{(\sapcta$\!\!$)}}.

\newparagraph
Aside from these data challenges, data engineering in enterprises comes with additional difficulties:

\newparagraph
\emph{(1) Task complexity:}$\:$
Enterprise tasks are often more complex than their academic formulations~\cite{savelieva_need_2022}.
One example is the task of entity matching, where existing literature often assumes that each entity is one row in a single table~\cite{konda_magellan_2016,peeters_using_2023}.
In enterprises, however, entities are often business objects that span across multiple tables, making matching difficult.
Moreover, the tasks themselves are often more complex.
For example, increasing the complexity of column type annotation by scaling to the true number of semantic types (5,089 instead of 200) in the SAP system leads to a further decrease in F1 scores, as shown in Figure~\ref{fig:headline_exp} \emph{(+ Task Challenges)}.

\newparagraph
\emph{(2) Internal knowledge:}$\:$
Data engineering in enterprises also often requires internal knowledge that is absent from public sources, limiting LLMs' abilities to understand the data without additional context. This is especially true for schema customizations that involve customer-defined semantic types. On SAP’s column type annotation dataset, when adding those columns to the task, F1 scores for such customer-defined columns are near zero, as shown in Figure~\ref{fig:headline_exp} \emph{(+ Knowledge Challenges)}.

\newparagraphwithindent
We want to highlight how such enterprise-specific challenges affect LLMs.
Our findings go beyond observations already known to the community and help unmask the jagged out-of-the-box capabilities of LLMs. 
For example, the impact of the complex enterprise schemas and symbolic values in the enterprise data is particularly striking. 
Moreover, we were surprised by how strongly the models resisted overwriting their parametric knowledge despite explicit guidance from documentation and examples.
We believe this to be the first attempt to examine LLMs for data engineering on real enterprise data at this breadth.
We see it as an important first step (of many) to make LLMs viable for enterprise data engineering.

\paragraph{Contributions.}
Our main contributions are:\footnote{This paper extends our previously published work \cite{bodensohn_llms_2024,bodensohn_automating_2024} on this topic.}
(1)~We systematically analyze the challenges involved in enterprise data engineering and structure them along the dimensions of enterprise \emph{data}, enterprise \emph{tasks}, and enterprise \emph{knowledge}.
(2)~We experiment on representative enterprise data to show how it differs from existing public benchmarks and understand how it affects LLMs.
(3)~We conduct multiple case studies that reflect real-world enterprise scenarios, allowing us to evaluate how different challenges affect LLMs in isolation.
(4)~We discuss directions for addressing these challenges, as well as the costs of using LLMs at enterprise scale.
(5)~Our experiments are performed with five recent LLMs from OpenAI, Anthropic, and Meta.
To enable follow-up research, we make the code---including the full evaluation setup, all prompts, and where legally possible also the data---of this paper available to the broader research community.

%% file: sections_02_background.tex
\begin{table}
  \caption{Model characteristics. We evaluate five LLMs from three model providers covering multiple types and sizes.}
  \label{tab:models}
  \centering
  \begin{threeparttable}
  \setlength{\tabcolsep}{5.5pt}
  \notesize{
  \begin{tabular}{l c c *2{S[table-format=2.2]}}
\toprule
&\multirow{2}{*}{\thead[c]{Context\\Window}}&\multirow{2}{*}{\thead[c]{Reason-\\ing}}&\multicolumn{2}{c}{\textbf{USD Per 1M }}\\
\cmidrule(lr){4-5}
&&&\textbf{Input}&\textbf{Output}\\
\midrule
\textbf{GPT-4o-Mini} (2024-07-18)\tnote{1}&128K&no&0.15&0.60\\
\textbf{GPT-4o} (2024-08-06)\tnote{1}&128K&no&2.50&10.00\\
\textbf{o1} (2024-12-17)\tnote{1}&200K&yes&15.00&60.00\\
\textbf{Claude~3.5~Sonnet}  (v1/v2)\tnote{2}&200K&no&3.00&15.00\\
\textbf{Llama~3.1~Instruct} (70B)\tnote{3}&128K&no&0.72&0.72\\
\bottomrule
\end{tabular}
}
\end{threeparttable}
\footnotesize{
     \begin{tablenotes}
     \hfill Pricing by OpenAI,$\!^1$ Anthropic,$\!^2$ and AWS Bedrock$^3$ in February 2025.
     \end{tablenotes}
    }
\end{table}

\begin{figure*}
  \centering
  \includegraphics[width=\linewidth]{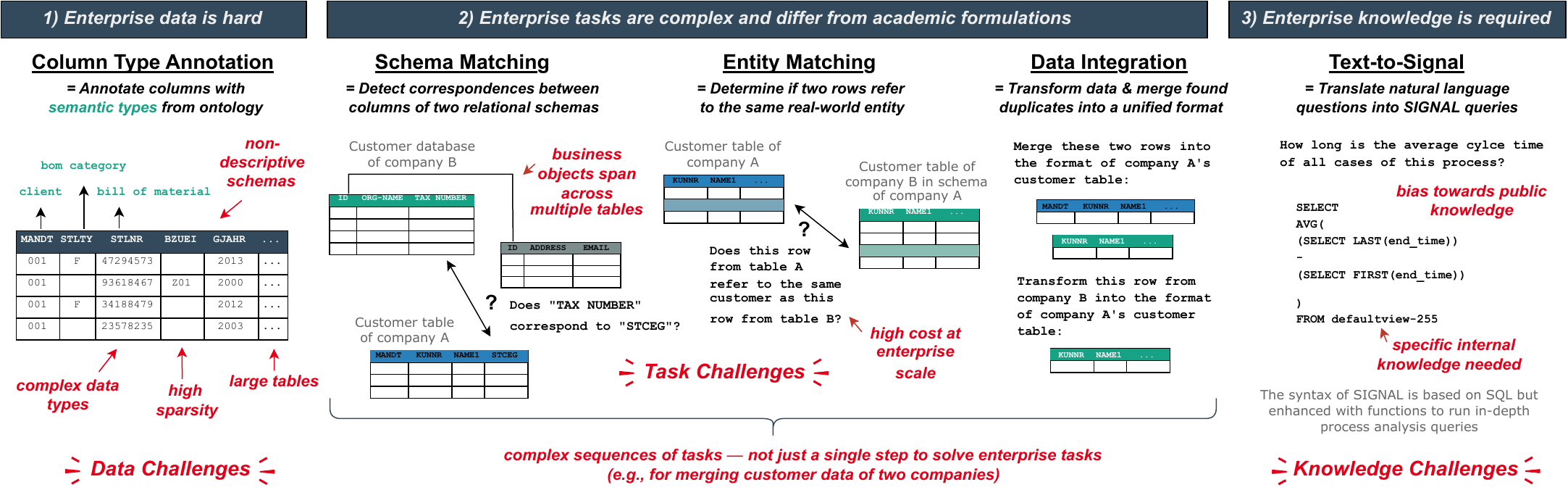}
  \caption{Enterprise-specific challenges in data engineering tasks. We use five well-established tasks serving as examples to highlight the breadth of challenges in enterprise settings and show their effect on LLMs for data engineering. The challenges shown here (e.g., high data sparsity) are highly general and extend to many other tasks beyond the ones shown here.}
  \label{fig:motivation}
\end{figure*}

\section{Setting of our Study}
\label{sec:background}

Using LLMs to solve table-based tasks is a promising research direction that has been actively studied in recent years~\cite{narayan_can_2022,zhang24data_preprocessors,peeters_using_2023,korini_column_2023,zhang_directions_2024}.
This section introduces the models we evaluate and briefly summarizes existing research on data engineering with LLMs.

\paragraph{Choosing LLMs for the study.}
LLMs are text foundation models trained on large corpora to complete natural language inputs~\cite{brown_language_2020} and follow user instructions~\cite{wei_2022_instruction_tuning,liu_pre-train_2023}.
Recent LLMs like GPT-4~\cite{openai_gpt-4_2024} and Llama~3~\cite{grattafiori_llama_2024} build on a range of further ideas, from mixture-of-experts architectures~\cite{shazeer2017outrageously} to more efficient implementations~\cite{dao_flashattention_2022}, leading to improvements in abilities like applying background knowledge and handling long inputs.
With the release of reasoning LLMs like OpenAI's o1~\cite{openai_2024_o1_blogpost} and Deepseek's R1 \cite{deepseekai_2025_r1}, they perform even better on tasks requiring several intermediate steps.

While many LLMs would be relevant candidates to include in our evaluation, our selection is necessarily limited.
To ensure generalizable results, we evaluate five recent LLMs from three model providers covering multiple types and sizes and including open and closed models.
As shown in Table~\ref{tab:models}, we use GPT-4o~\cite{openai_gpt-4o_2024} as the state-of-the-art and \mbox{GPT-4o-Mini} as the cheapest model from OpenAI in February 2025, and \mbox{o1}~\cite{openai_openai_2024} as a reasoning model.
We further include Claude~3.5~Sonnet~\cite{anthropic_claude_2024} from Anthropic and Llama~3.1~Instruct~\cite{meta_llama_2024} from Meta.
In general, LLMs bear the challenge of output variability. 
In all our experiments, we reduce randomness to a minimum by setting the sampling temperature to 0.
We also repeat several of our experiments multiple times, witnessing only small standard errors and no change in the overall characteristics of the results.

\paragraph{Positioning the study.}
Although several papers have already observed the differences between enterprise and web data~\cite{vogelsgesang_get_2018,savelieva_need_2022,kang_schema_2003,zhu_josie_2019}, existing research on data engineering with LLMs primarily uses evaluation datasets based on tables from public web resources, calling the applicability of LLMs on real-world enterprise data into question.
More recently, some enterprise-specific benchmarks have been released for tabular prediction tasks~\cite{klein_salt_2024} and Text-to-SQL~\cite{papicchio2025qatch, chen_beaver_2024}.
Related to our work, \citet{kayali_mind_2025} quantify performance gaps between private and public data for the task of column type annotation and find that benchmarks based on public data overestimate the performance of LLMs.
However, these studies focus only on individual tasks and consider the data as the only challenge in enterprise scenarios.
In contrast, we perform case studies on various tasks to systematically analyze the challenges along several dimensions of working in enterprise scenarios.

\paragraph{Working with real enterprise data.}
Attempts like ours often fail because enterprise data is usually highly confidential and, therefore, hard to use in evaluations.
For this paper, we were able to experiment with actual customer data from the enterprise systems of SAP.
While one could argue that this study is still limited because we only use data from SAP, and there are clearly many alternative enterprise systems, SAP stands out as a dominant player in enterprise software systems across multiple industries worldwide.
As such, we believe that our insights based on SAP data are highly valuable on their own and hope that our paper inspires other researchers with access to similar enterprise datasets to repeat our evaluations on their data.
Importantly, the core characteristics of SAP data reflect the findings from other papers observing the differences between enterprise and web data \cite{vogelsgesang_get_2018,savelieva_need_2022,kang_schema_2003,zhu_josie_2019}, highlighting the generality of our results beyond SAP data.

%% file: sections_03_method.tex
\section{Design of our Study}
\label{sec:method}

\begin{table*}
  \caption{Data characteristics of publicly available benchmark datasets compared to representative customer data from SAP. Enterprise tables have substantially more rows and columns and display a higher sparsity compared to public data. Although most attributes are of type \texttt{NVARCHAR}, the data is highly symbolic, and table names, column names, and cell values are often not human readable because of abbreviations and enterprise-specific encodings.}
  \label{tab:data_characteristics}
  \centering
\begin{threeparttable}
  \setlength{\tabcolsep}{8pt}
  \notesize{
  \centering
  \begin{tabular}{l rrrrrcccrr}
\toprule
&\multirow{2}{*}{\textbf{Tables}}&\multicolumn{2}{c}{\textbf{Columns}}&\multicolumn{2}{c}{\textbf{Rows}}&\multirow{2}{*}{\textbf{Sparsity\tnote{1}}}&\multicolumn{2}{c}{\textbf{Data Types\tnote{2}}}&\multicolumn{2}{c}{\textbf{Column Type Annotation}}\\
\cmidrule(lr){3-4}
\cmidrule(lr){5-6}
\cmidrule(lr){8-9}
\cmidrule(lr){10-11}
&&\textbf{Med}&\textbf{95th}&\textbf{Med}&\textbf{95th}&&\textbf{\textit{abc}}&\textbf{\textit{123}}&\textbf{Column Types}&\textbf{Labeled Columns}\\
\midrule
\textbf{WikiTables-TURL}&397,098&1&3&8&43&0.12&1.00&0.00&255&628,254\\
\textbf{SOTAB}&59,548&7&17&33&721&0.08&0.85&0.15&91&162,351\\
\textbf{GitTablesCTA}&1,100&12&33&25&263&0.12&0.33&0.67&122 | 59\tnote{3}
&2,517 | 1,374\tnote{3}\\
\textbf{SportsTables}&1,183&21&31&32&924&0.07&0.16&0.84&452&24,821\\
\midrule
\textbf{\sapcta}&100\tnote{4}&46&343&473,038&50,836,964&0.43&0.55&0.45\tnote{5}&5,089&8,106\\
\bottomrule
\end{tabular}
}
    \footnotesize{
     \begin{tablenotes}
        \item[1] Sparsity is the fraction of empty cells.
        \item[2] Non-numerical \textit{(abc)} and numerical \textit{(123)} columns determined by pandas.
        \item[3] Using semantic types from DBPedia | Schema.org.
        \item[4] We experiment on a representative sample from the thousands of tables in the customer system.
        \item[5] Only 14\% of columns have numerical SQL types like \texttt{INT} and \texttt{DECIMAL}.
     \end{tablenotes}
    }
  \end{threeparttable}
\end{table*}

We aim to take a holistic view of data engineering in enterprises by systematically analyzing the out-of-the-box performance of LLMs, covering enterprise-specific challenges along the dimensions of \emph{data}, \emph{tasks}, and \emph{knowledge}.
Figure \ref{fig:motivation} provides an overview of the tasks we have chosen to analyze the performance along each dimension.
Below, we explain the rationale for selecting these tasks as examples to examine challenges that arise broadly across different enterprise tasks and scenarios.
Sections \ref{sec:data} to \ref{sec:knowledge} then describe our experimental evaluation for each of these challenges.
  
\paragraph{The data challenge.} 
Enterprise data differs from public evaluation datasets in various aspects.
First, the tables in enterprise databases are substantially larger in their number of columns and rows, and the schema and data are often more complex with column names and values that lack intuitive meaning.
Moreover, enterprise data exhibits a much higher sparsity than public data, leaving many cells empty.
Together, these factors make data engineering on enterprise data much more challenging.
To explore these challenges, we focus on the task of column type annotation (Figure~\ref{fig:motivation} left)~\cite{hulsebos_sherlock_2019,zhang_sato_2020}.
We have chosen this task because it is well-studied in the literature and thus enables the comparison of accuracies with public datasets.
Moreover, the task can be formulated in various ways, stressing different data factors (e.g., with and without schema information) and allowing us to study the effects of different challenges like table size, sparsity, and descriptiveness of the data and schema.

\paragraph{The task challenge.} 
Beyond the complexities of enterprise data, the enterprise tasks themselves are also more complex. 
First, while academia typically studies data engineering tasks in isolation, enterprise tasks in practice are often compounds of multiple simpler data engineering steps.
As such, we argue that the evaluation procedures themselves must change and we need to study accuracy end-to-end to analyze effects such as how errors propagate.
In this paper, we examine the compound nature of enterprise tasks using the example of integrating two customer databases, as illustrated in Figure~\ref{fig:motivation} (middle).
This process involves the steps of schema matching, entity matching, and data integration. 
A second challenge for enterprise tasks (not shown in Figure~\ref{fig:motivation}) is that even the individual steps are often more complex.
For example, while entity matching in academia assumes 1:1 matches across rows of two tables \cite{peeters_wdcproducts_2024, mudgal2018deep}, matching in enterprise scenarios covers more complex cases like mapping several bank transfers to one invoice, and the required data is often scattered across multiple tables.

\paragraph{The knowledge challenge.}
As a last challenge, data engineering in enterprises often requires enterprise-specific knowledge.
This is particularly challenging because LLMs will likely not have seen the required information during training, as it is covered only in internal documentation or even just certain implementation details in the code of enterprise systems.
To analyze these challenges, we have selected a task related to data exploration that requires translating natural language queries about business processes into the enterprise-specific query language SIGNAL, which differs slightly from SQL (Figure~\ref{fig:motivation} right)~\cite{kampik_signal_2023}.
The task is interesting because LLMs should, judging by its similarity to Text-to-SQL, in principle be able to solve it.
However, they lack particular information about how exactly SIGNAL differs from SQL.
At the core, the question we want to answer here is to what extent the lack of enterprise knowledge affects the accuracy of LLMs, and whether it can be increased by providing enterprise-specific knowledge as context.  

\paragraph{An orthogonal challenge: cost.} Since enterprise tables can contain millions of rows, using LLMs on such data can cause high costs, rendering some of the larger, more complex LLMs economically unviable.
We discuss this aspect in Section~\ref{sec:costs}.

%% file: sections_04_data.tex
\section{The Data Challenge}
\label{sec:data}
In this section, we quantify the anatomy of enterprise data by comparing real-world customer data from SAP to publicly available table corpora. 
We point out four challenges specific to enterprise data and perform experiments to evaluate how they affect LLMs, using column type annotation as an example task (Figure~\ref{fig:motivation} left).

\subsection{Data Challenges}
\label{sec:data_challenges}

For our study, we constructed a new corpus called \sapcta$\!$.
Table~\ref{tab:data_characteristics} compares the data characteristics of the real-world customer data in \sapcta to several publicly available column type annotation benchmarks: WikiTables-TURL~\cite{deng_turl_2020}, SOTAB~\cite{korini_sotab_2022}, GitTablesCTA~\cite{madelon_hulsebos_2021_5706316}, and SportsTables~\cite{langenecker_sportstables_2023}.
We observe the following differences:

\challenge{C1: Table size.}
A first important observation is that enterprise tables often have substantially more rows and columns than the tables in public corpora.
As shown in Table~\ref{tab:data_characteristics}, some tables have hundreds of columns and millions of rows.
While the large scale is a well-established data management problem~\cite{zhu_josie_2019}, it poses challenges for LLMs, which have limited context windows.
Although recent models have extended context windows, feeding large tables into LLMs still has downsides since latency and cost depend on the input size, and recent studies have shown that long contexts can lead to degraded performance for data residing in the ``middle''  \cite{liu_lost_2023}.

\challenge{C2: Descriptiveness.}
Another important insight is that schema properties like table and column names are often not descriptive but rather abbreviations that can only be understood with background knowledge or additional metadata~\cite{kang_schema_2003}.
This additional metadata is often unavailable or may not fit into the context window of the LLM.
Moreover, the background knowledge is often specific to the particular enterprise, causing challenges for LLMs trained exclusively on publicly available data, as we discuss in Section~\ref{sec:knowledge}.

\challenge{C3: Sparsity.}
A third insight is that enterprise data is highly sparse.  Table~\ref{tab:data_characteristics} shows that on average, 43\% of the cells in enterprise tables are empty, compared to only 7-12\% in existing datasets.
The high sparsity results in a significant lack of information, posing a challenge for LLMs which rely on contextual cues to make accurate predictions.
Moreover, we find that in addition to empty values, the cells in enterprise tables often contain dummy values such as \texttt{00000} that also denote the absence of an actual value.

\challenge{C4: Data types.}
Interestingly, we find that only 45\% of the columns in \sapcta are classified as numerical by pandas, suggesting that enterprise datasets can be more text-heavy than found in previous studies \cite{langenecker_sportstables_2023}.
Moreover, only 14\% of the columns in the database schema have numerical data types like \texttt{INT} and \texttt{DECIMAL}.
A closer inspection of the actual data reveals that the non-numerical data type \texttt{NVARCHAR} is often used to store symbolic values and codes such as invoice and material numbers, which is in line with previous findings~\cite{vogelsgesang_get_2018}.
Since these values are not self-expressive, LLMs cannot make use of them without additional context.

\subsection{Experiments \& Results}
\label{sec:data_experiments_results}

To study how the challenges C1-C4 affect LLMs for data engineering, we compare the models' performance on our real-world enterprise dataset \sapcta to their performance on existing evaluation datasets.

\subsubsection*{Task.}
We evaluate on column type annotation, a well-established data engineering task where the goal is to annotate the columns of a relational table with semantic types from a pre-defined ontology~\cite{hulsebos_sherlock_2019,zhang_sato_2020}. 
We see it as an interesting example task to uncover the challenges of understanding enterprise tables with LLMs, since it requires a semantic understanding of the content of each column as well as the values of other columns and the table schema, all of which provide important signals to derive a semantic type.

\subsubsection*{Setup.}
Our \sapcta corpus spans diverse business domains such as Finance, Sales and Distribution, Material Management, and Production Planning.
For our experiments, we select 100 representative tables from the larger corpus, which contains multiple thousands of tables. The selection is based on discussions with SAP experts to identify the most widely used tables in the system.
In total, the dataset includes 5,089 semantic types, such as \textit{Amount Difference in Local Currency} or \textit{Product Cost Collector}, which are generally more fine-granular than those used in other datasets \cite{korini_sotab_2022, madelon_hulsebos_2021_5706316}.

Our prompting strategy builds on best practices from existing literature, where the model annotates the columns of a given table based on a list of semantic types and one randomly-selected example included in the prompt~\cite{zhang_sato_2020, korini_column_2023}.$\!$\footnote{To focus solely on the data challenges, we include only a subset of the semantic types in every prompt similar to the setup from Figure~\ref{fig:headline_exp} \emph{(\sapcta$\!\!$)}.}
Based on ablations (see Appendix~\ref{appendix:column_type_annotation} in our extended technical report), we limit each table to three randomly-selected rows as a reasonable trade-off between performance and cost.
For similar reasons, we serialize the tables in CSV format, which requires fewer formatting tokens than other serialization schemes like Markdown and JSON \cite{singha_tabular_2023, sui_table_2024}.
We instruct the model to generate the column types as a JSON-formatted list.

\experiment{Exp. 1: Enterprise vs. public tables.}
In our first experiment, we compare the performance for column type annotation on our \sapcta corpus with the performance on GitTablesCTA \cite{madelon_hulsebos_2021_5706316} and SportsTables \cite{langenecker_sportstables_2023}.
We perform each experiment twice, with and without including the table and column names (i.e., the table schema) in the prompt.
Existing evaluations typically leave out the column names, since the semantic types are directly derived from them and the task would thus become trivial.
For \sapcta$\!$, however, the task is much harder.
Therefore, we want to investigate how much the additional information helps.

Table \ref{tab:data_cta_headers} shows the results of this experiment. 
We make the following observations: 
First, LLMs have severe issues with column type annotation on enterprise data, leading to substantially worse results compared to the public benchmarks GitTablesCTA and SportsTables.
The results are particularly poor in the experiments without table and column names, 
indicating that enterprise data on its own contains few helpful signals. 
Adding table and column names to the prompt improves the results, but they still remain much lower than for web tables.
The remaining performance gap could potentially be attributed to the non-descriptive schema, the extremely wide and sparse tables, and the complex data types.

\begin{table}
  \caption{Enterprise vs. public tables. The table shows support-weighted F1 scores for column type annotation with and without column names. The results on enterprise data are substantially worse than on existing benchmarks.}
  \label{tab:data_cta_headers}
  \centering
  \setlength{\tabcolsep}{5pt}
  \notesize{
  \begin{tabular}{l *6{S[table-format=1.2]}}
\toprule
&\multicolumn{2}{c}{\textbf{GitTablesCTA}}
&\multicolumn{2}{c}{\textbf{SportsTables}}
&\multicolumn{2}{c}{\textbf{\sapcta}}
\\
\cmidrule(lr){2-3} \cmidrule(lr){4-5} \cmidrule(lr){6-7}
\textbf{Column Names}&\textbf{w/out}&\textbf{with}&\textbf{w/out}&\textbf{with}&\textbf{w/out}&\textbf{with}\\
\midrule
\textbf{GPT-4o-Mini} & 0.52 & 0.96 & 0.27 & 0.55 & 0.02 & 0.07 \\
\textbf{GPT-4o} & 0.56 & \textbf{0.99} & 0.57 & 0.91 & 0.04 & 0.24 \\
\textbf{Claude 3.5 Sonnet} & \textbf{0.67} & 0.96 & \textbf{0.66} & \textbf{0.93} & \textbf{0.05} & \textbf{0.34} \\
\textbf{Llama 3.1 Instruct} & 0.49 & 0.95 & 0.38 & 0.73 & 0.02 & 0.10 \\
\bottomrule
\end{tabular}
}
\end{table}

\experiment{Exp. 2: Textual vs. numerical data.}
LLMs are known to often perform better on textual data than on numerical data \cite{frieder_mathematical_2023, langenecker_sportstables_2023}. 
To study this effect on enterprise data, we compare the performance for non-numerical and numerical columns in our \sapcta corpus.

Table \ref{tab:data_cta_data_types} shows that, as expected, we see a higher performance on non-numerical enterprise data.
By contrast, the results on the public benchmarks GitTablesCTA and SportsTables remain inconclusive, with only small performance gaps between non-numerical and numerical columns that also differ between models.
The low performance on the \sapcta dataset indicates that numerical data in enterprise systems is even harder to understand than in public benchmarks.
Furthermore, the low scores for non-numerical columns may stem from the fact that enterprise tables often store identifiers like \texttt{INV0014056} as type \texttt{NVARCHAR}.

\experiment{Exp. 3: Table width and sparsity.}
To further investigate the performance gap between public benchmarks and enterprise data, we incrementally adapt the enterprise tables to resemble the characteristics of web tables more closely.
Since two of the main differences are table width and sparsity, we vary the number of columns per table by randomly sampling subsets of columns and vary the sparsity by initially selecting only non-sparse columns and randomly removing individual cell values.

\begin{figure}
    \includegraphics[width=\linewidth]{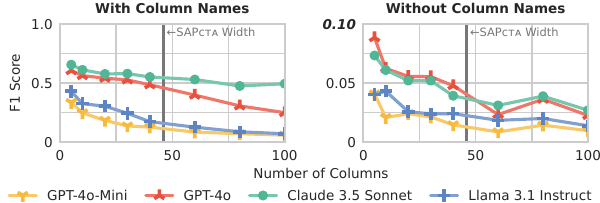}
    \caption{Effect of the table widths. The plots show support-weighted F1 scores for column type annotation with and without column names (zoomed in on the right). Increased table widths lead to worse results.}
    \label{fig:data_width}
\end{figure}

\begin{figure}
    \includegraphics[width=\linewidth]{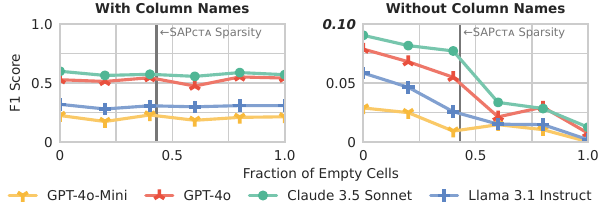}
    \caption{Effect of the sparsity. The plots show support-weighted F1 scores for column type annotation with and without column names (zoomed in on the right). Increased sparsity leads to worse results if no column names are given.}
    \label{fig:data_sparsity}
\end{figure}

Figure~\ref{fig:data_width} shows that increasing table widths lead to substantially worse results, indicating that the large table widths in enterprise data are indeed a major problem for LLMs.
Figure~\ref{fig:data_sparsity} shows that increased sparsity leads to worse results only if no table and column names are provided, whereas with table and column names, increased sparsity does not change the results much.
This indicates that on enterprise data, LLMs rely primarily on the column names to predict the semantic types, maintaining almost constant accuracy even if no cell values are provided (Figure~\ref{fig:data_sparsity} left, sparsity of 1.0).

\begin{table}
  \caption{Non-numerical \textit{(abc)} vs. numerical \textit{(123)} data. The table shows support-weighted F1 scores for column type annotation with column names. The results on numerical enterprise data are consistently worse.}
  \label{tab:data_cta_data_types}
  \centering
  \setlength{\tabcolsep}{6.2pt}
  \notesize{
  \begin{tabular}{l *6{S[table-format=1.2]}}
\toprule
&\multicolumn{2}{c}{\textbf{GitTablesCTA}}
&\multicolumn{2}{c}{\textbf{SportsTables}}
&\multicolumn{2}{c}{\textbf{\sapcta}}
\\
\cmidrule(lr){2-3} \cmidrule(lr){4-5} \cmidrule(lr){6-7}
\textbf{Data Types}&\textbf{\textit{abc}}&\textbf{\textit{123}}&\textbf{\textit{abc}}&\textbf{\textit{123}}&\textbf{\textit{abc}}&\textbf{\textit{123}}\\
\midrule
\textbf{GPT-4o-Mini} & 0.97 & 0.95 & 0.68 & 0.53 & 0.11 & 0.03 \\
\textbf{GPT-4o} & \textbf{0.99} & \textbf{0.98} & \textbf{0.87} & 0.91 & 0.31 & 0.16 \\
\textbf{Claude 3.5 Sonnet} & 0.97 & 0.95 & 0.81 & \textbf{0.96} & \textbf{0.41} & \textbf{0.27} \\
\textbf{Llama 3.1 Instruct} & 0.94 & 0.96 & 0.85 & 0.72 & 0.15 & 0.05 \\
\bottomrule
\end{tabular}
}
\end{table}

\experiment{Exp. 4: Improvement strategies.}
Figure~\ref{fig:data_width} shows that column type annotation accuracy decrases as the table width increases.
To mitigate this effect, we explore three strategies for handling wide tables:
\emph{All columns (full table)} is our baseline setup, where the model predicts all semantic types of the full table in a single call.
\emph{Chunks of $N$ columns} divides the table into chunks of $N \in \{50, 10, 1\}$ columns and prompts the model for each chunk separately without the full table as context.
\emph{One column at a time (with full table context)} predicts each column type in a separate LLM call, but provides the full table in the prompt. We state the column to be predicted either \emph{by name} or \emph{by index} to analyze which variant works better.

Table~\ref{tab:data_cta_headers_strategies} shows that on public benchmarks, the differences between strategies are small, although smaller chunk sizes reduce performance if no column names are provided.
This indicates that the data in public benchmarks provides strong contextual signals for column type annotation.
For \sapcta{}$\!$, however, chunking without column names does not further degrade the already low performance.
With column names, chunking even improves results.
Notably, for \sapcta{}$\!$, the one-column-at-a-time strategy using the full table as context yields the best results.
Since enterprise data without headers provides little signal, we suspect that the models rely primarily on the column names, largely ignoring the actual data.
This method is also the most expensive, requiring one LLM call per column with the full table as context.
Inference over the full \sapcta dataset with GPT-4o costs USD 53.85, compared to USD 4.37 for the baseline—an order of magnitude more.
Yet even with this setup, performance remains far below that on public benchmarks.

\subsection{Discussion}
\label{sec:data-discussion}

\subsubsection*{Learnings.}
Our analysis of the customer data from SAP shows that enterprise data differs fundamentally from public benchmarks, and our experiments demonstrate that these differences lead to substantial performance decreases.
The performance declines with higher sparsities and larger table widths, indicating that LLMs are unreliable at scale and sometimes fail to uphold even the basic table structure by generating an incorrect number of labels.
Moreover, the non-descriptive schemas and complex data types pose significant challenges for LLMs.
While we have shown this for the task of column type annotation, these data challenges will appear again throughout the following sections for other data engineering tasks, highlighting that they are general and affect a broad range of tasks.

\subsubsection*{Outlook.}
We believe that overcoming these challenges requires better representations for enterprise data.
A first promising direction is the development of foundation models for relational data~\cite{vogel_towards_2022,wu_learning_2025}.
While there is already a large body of research on foundation models for tabular data~\cite{qu_tabicl_2025,he_tablelora_2025,li_table-gpt_2024,yang_tableformer_2022,eisenschlos_mate_2021,herzig_tapas_2020,yin_tabert_2020,deng_turl_2020}, computing representations for complex enterprise data is still an open research problem.
Some LLM weaknesses like their low reliability at scale require rethinking how they are applied to enterprise data, for example by using chunking to handle large tables.
One way to alleviate the low self-expressiveness of the enterprise data could be to better contextualize the raw data by incorporating metadata such as data dictionaries, which contain textual descriptions for table and column names and for symbolic values.

\begin{table}
  \caption{Strategies for handling wide tables, including our baseline setup predicting \emph{all columns} in one call, \emph{chunking}, and predicting just \emph{one column with full table context} per call. The table shows support-weighted F1 scores for column type annotation with and without column names.}
  \label{tab:data_cta_headers_strategies}
  \centering
  \setlength{\tabcolsep}{2pt}
  \footnotesize{
  \begin{tabular}{p{0.2cm}l *6{S[table-format=1.2]}}
\toprule
&&\multicolumn{2}{c}{\textbf{GitTablesCTA}}
&\multicolumn{2}{c}{\textbf{SportsTables}}
&\multicolumn{2}{c}{\textbf{\sapcta}}
\\
\cmidrule(lr){3-4} \cmidrule(lr){5-6} \cmidrule(lr){7-8}
\multicolumn{2}{l}{\textbf{Column Names}}&\textbf{w/out}&\textbf{with}&\textbf{w/out}&\textbf{with}&\textbf{w/out}&\textbf{with}\\
\midrule
\multirow{6}{*}{\rotatebox{90}{\textbf{GPT-4o-Mini$\quad\,$}}} & \textbf{All columns (full table)} & \textbf{0.52} & 0.96 & 0.27 & 0.55 & 0.02 & 0.07 \\
\cmidrule(){2-8}
&\textbf{Chunks of 50 columns} & \textbf{0.52} & 0.96 & \textbf{0.29} & 0.51 & 0.01 & 0.09 \\
&\textbf{Chunks of 10 columns} & 0.48 & \textbf{0.97} & 0.20 & \textbf{0.62} & 0.01 & 0.22 \\
&\textbf{Chunks of 1 columns} & 0.22 & 0.92 & 0.06 & 0.61 & 0.01 & 0.20 \\
\cmidrule(){2-8}
&\textbf{1 column by index (full table)} & 0.30 & 0.66 & 0.07 & 0.11 & \textbf{0.03} & 0.11 \\
&\textbf{1 column by name (full table)} & \text{-} & 0.89 & \text{-} & 0.50 & \text{-} & \textbf{0.32} \\
\midrule
\multirow{6}{*}{\rotatebox{90}{\textbf{GPT-4o$\quad\;\,$}}} & \textbf{All columns (full table)} & \textbf{0.56} & \textbf{0.99} & 0.57 & \textbf{0.91} & 0.03 & 0.27 \\
\cmidrule(){2-8}
&\textbf{Chunks of 50 columns} & 0.54 & \textbf{0.99} & \textbf{0.58} & 0.89 & 0.03 & 0.44 \\
&\textbf{Chunks of 10 columns} & 0.51 & \textbf{0.99} & 0.45 & \textbf{0.91} & 0.03 & 0.46 \\
&\textbf{Chunks of 1 columns} & 0.29 & 0.90 & 0.11 & 0.83 & 0.02 & 0.44 \\
\cmidrule(){2-8}
&\textbf{1 column by index (full table)} & 0.37 & 0.85 & 0.23 & 0.36 & \textbf{0.05} & 0.26 \\
&\textbf{1 column by name (full table)} & \text{-} & 0.86 & \text{-} & 0.89 & \text{-} & \textbf{0.52} \\
\bottomrule
\end{tabular}
}
\end{table}

%% file: sections_05_tasks.tex
\section{The Task Challenge}
\label{sec:tasks}

A second dimension where data engineering in enterprise settings stands out is the complexity of the tasks.
The academic definition of a data engineering task often makes simplifying assumptions that do not hold in real-world scenarios.
For example, literature on entity matching often assumes that we compare the similarity of individual rows of two tables~\cite{peeters_using_2023,konda_magellan_2016}.
By contrast, business entities in enterprises often span across multiple tables, forming graphs of tuples.
At the same time, enterprises approach data engineering with broad business objectives in mind.
Therefore, enterprise tasks are often composed of multiple steps (e.g., combining schema matching and entity matching). 
In this section, we discuss how enterprise tasks differ in their nature from the tasks evaluated in academia and examine how these differences affect LLMs.

\subsection{Task Challenges}
\label{sec:task_challenges}

We base our analysis of the task-related challenges in enterprise data engineering on the following two case studies:

\subsubsection*{Case Study 1: Enterprise entity matching.} 
\begin{figure}
    \includegraphics[width=\linewidth]{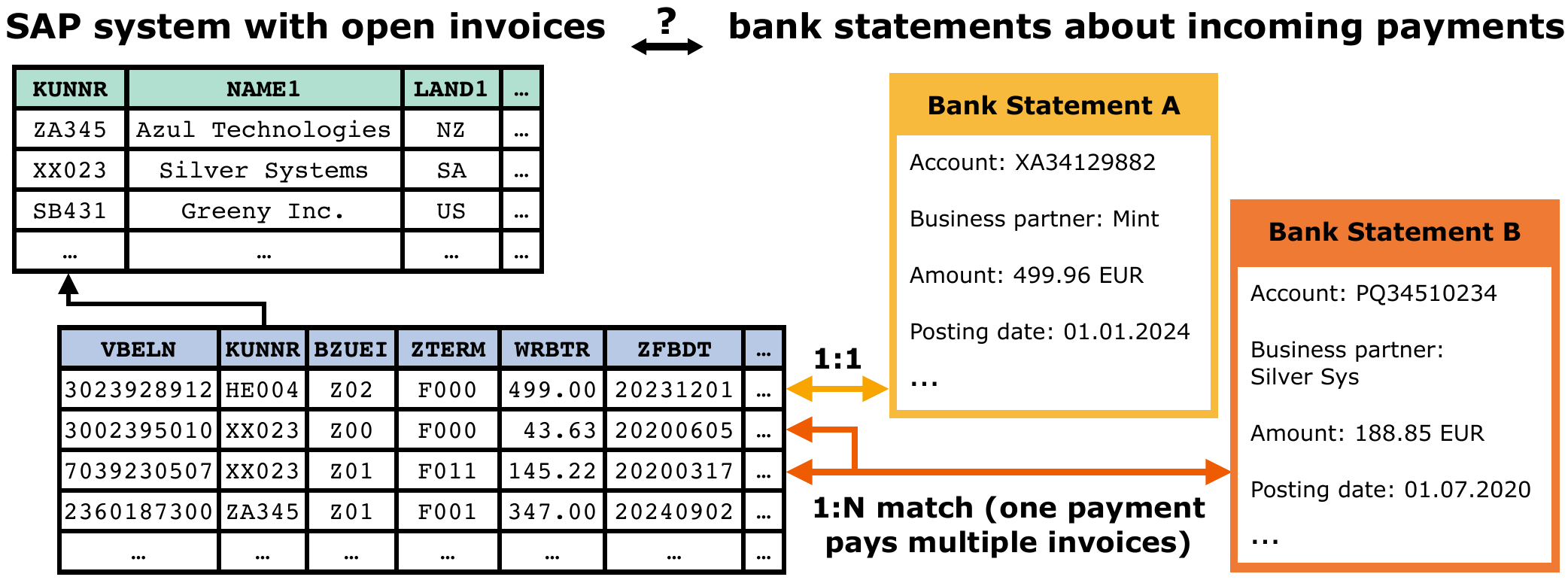}
    \caption{Enterprise entity matching: Bank statements of incoming payments must be matched to open invoices. Challenges include the invoices being represented by multiple tables, multi-match cases where a single payment pays multiple invoices, and discrepancies in amounts and descriptions.}
    \label{fig:pay_to_inv}
\end{figure}

Entity matching is a data engineering task that often occurs in enterprise scenarios.
We study the scenario shown in Figure~\ref{fig:pay_to_inv}, in which the bank statements of incoming payments must be matched to a company's open invoices.
The scenario showcases many challenges of entity matching in enterprises, including problems like complex table structures and the fact that matching requires finding 1:N or even N:M matches.

\subsubsection*{Case Study 2: Enterprise database integration.} 
For our second case study, we look into another enterprise scenario where two customer databases from different companies must be merged.
To ensure a unified and consistent customer database, this process involves the individual steps of schema matching, customer record matching, and data integration, as illustrated in Figure \ref{fig:motivation} (center).

\newparagraph
In the following, we highlight the general challenges that arise in enterprise tasks and provide examples based on the two scenarios.

\challenge{C5: Entities span multiple tables.}
Entities in enterprise systems are often business objects represented by multiple rows stored in different connected tables.
In SAP systems, data pertaining to a particular material is scattered across the \texttt{MARA}~(material type and basic statistics), \texttt{MARC}~(manufacturing-related details), \texttt{MBEW}~(valuation data), and other tables.
The same holds true for the entities in our first case study \emph{(entity matching)}.
For many data engineering tasks, one must therefore either manually construct views that extract the relevant fields into a single table, or approaches must work directly with the complex table structures that form a business entity.

\challenge{C6: Compound tasks.}
Whereas data engineering in research is often addressed as isolated problems, such as deduplication~\cite{papenbrock_progressive_2015} and missing value imputation~\cite{mei_capturing_2021}, tasks in enterprise contexts are typically approached on a more holistic level concerning broader business objectives.
Instead of focusing on individual tasks, enterprises aim to solve end-to-end workflows, as highlighted by our second case study \emph{(database integration)}.
While the steps of such compound tasks could be executed sequentially, errors often propagate and amplify in later steps.
Analyzing each step in isolation does not reveal the overall quality of the task in enterprise scenarios.

\challenge{C7: Task complexity.}
Even individual workflow steps can be more complex than their counterparts studied in academia.
For example, the entities to be matched in public entity matching datasets are usually of the same type, such as e-commerce products, restaurants, and scholarly articles~\cite{konda_magellan_2016, peeters_wdcproducts_2024}. 
By contrast, enterprise scenarios often require matching between different types of entities, like products to commodity codes or payments to invoices, which have overlapping but different sets of attributes.
An additional challenge in enterprise scenarios is that the matches are often not 1:1 matches as in the literature, but can also be 1:N, N:1, or even N:M matches, making the problem much harder. 
In our payment-to-invoice matching scenario, it is common for a customer to pay multiple invoices with only one payment~(1:N), or for one invoice to be paid by multiple payments~(N:1), such as in the case of down payments.

\challenge{C8: Data complexity on top.}
Along with these task complexities, the complexity of the data makes the tasks themselves also harder.
For example, when matching payments to invoices, the memo lines of incoming bank statements are not standardized, but rather free-form text entered by customers.
Therefore, complex errors occur regularly, making automated matching difficult. Compared to such data errors in enterprise scenarios, the errors in academic entity matching datasets are often much simpler. 

\subsection{Experiments \& Results (Case Study 1)}
\label{sec:case_study_1}

To demonstrate how these challenges affect data engineering with LLMs, we first conduct experiments for our first case study \emph{(entity matching)}.
Experiments for our second case study \emph{(database integration)} follow in the next subsection.

%For our experiments, we are using a dataset based on real-world customer data, which we describe next.

\subsubsection*{Setup.}
For our first case study, we use a dataset of payments and invoices following the characteristics of the actual data from an enterprise system and reflecting the challenges described above~(see Appendix~\ref{appendix:case_study_1} in our extended technical report).
The dataset contains 15,521 invoices and 12,332 payments.
Following existing literature~\cite{narayan_can_2022,peeters_using_2023}, we formulate the entity matching task as a binary classification where the LLM decides if two entities (one payment and one invoice) match; i.e., if the payment pays for the invoice.
Note that some of these combinations are part of multi-match cases where one payment pays for multiple invoices or vice versa.
To keep the costs tractable, we use a subset of the dataset consisting of 790 payment-invoice combinations where the payment does pay for the invoice and 1,210 combinations where it does not.

For each combination, we prompt the LLM to predict if it is a match or a non-match. Furthermore, we experiment with presenting the invoices in two formats to the LLM: one format where each invoice is represented by multiple separate tables, and another format where each invoice is represented as one row in a flat view that combines all relevant columns from these tables, mimicking the traditional entity matching setting.
We use a few-shot approach that includes one positive and one negative example in the prompt, as zero-shot performance was low on enterprise data.
More examples did not improve the accuracy much.
Starting from this simple setup, we make the task incrementally more complex.

\experiment{Exp. 5: Increasing task complexity.} 
In the following, we explain the different scenarios with results provided in Table \ref{tab:tasks_pay_to_inv}:

\newparagraph
\emph{1. (Clean)}$\:$
First, we evaluate on clean 1:1 matches using the flat table views described above, which is closest to the academic setting.
In this setting, the payment's memo line includes the correct reference numbers, the payment amount is exactly as stated in the invoice, and the customer name is also identical.
As shown in Table~\ref{tab:tasks_pay_to_inv}~\emph{(Clean)}, all models reach very high F1 scores in this setup.
Next, we incrementally increase the difficulty based on the challenges we observe in the enterprise scenario.

\newparagraph
\emph{2. (+ Errors)}$\:$
As described in C8, data errors in enterprises can differ from those in academic benchmarks, making the matching harder. 
To demonstrate this, we add representative errors and discrepancies that typically occur in real-world customer data to the payments and invoices, such as missing or additional digits in the reference numbers or minor discrepancies in the paid amount.
As shown in Table~\ref{tab:tasks_pay_to_inv}~\emph{(+ Errors)}, these seemingly small inconsistencies already cause a noticeable drop in accuracy, highlighting the sensitivity of the models to such errors.
We further study this effect in Exp.~6, showcasing that LLMs have substantial problems understanding the semantics of enterprise data.

\newparagraph
\emph{3. (+ Multi-Matches)}$\:$
Next, we make the matching even harder by focusing on the multi-match cases described in C7, where one payment pays for multiple invoices or multiple payments together pay for one invoice, a setting that is already closer to the real-world scenarios found in enterprises.
As shown in Table~\ref{tab:tasks_pay_to_inv}~\emph{(+ Multi-Matches)}, this vastly increases the task's difficulty, since amounts are split between multiple payments and payments might include multiple reference numbers for different invoices.
To understand the root cause of the performance decrease and the high deviation between the models, we further investigate the precision and recall~(see Appendix~\ref{appendix:case_study_1} in our extended technical report).
We find that while all models achieve a very high precision close to $1.00$, the differences in F1 scores are primarily driven by variations in recall.
This indicates that all models take a rather cautious approach, predicting matches only when they are highly certain.
As a result, they prioritize high precision in matching and therefore miss many correct matches.
While this high precision is sometimes preferred over high recall but low precision, the high rate of missed matches causes substantial manual efforts for enterprises to find the missing matches.

\newparagraph
\emph{4. (+ Multiple Tables)}$\:$
Lastly, to show the impact of business entities spanning across multiple tables as described in C5, we represent invoices using multiple separate tables (as in the original SAP schema) instead of a single flat table view.
While metadata about each invoice is stored in one table, specific information like the amount and due date is stored in a second table, and information about customers is stored in another separate customer table.
In this scenario, which now closely resembles the actual enterprise task in its full complexity, all models see large performance decreases compared to the previous experiments, indicating that the models have difficulties working with the complex data structures used in enterprises.

\begin{figure}
    \includegraphics[width=\linewidth]{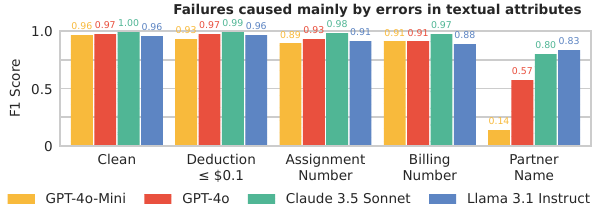}
    \caption{Impact of different error categories when matching payments to invoices. 
    Discrepancies in the partner name attribute cause a substantially larger drop than numerical attributes, indicating a strong reliance on textual fields.}
    \label{fig:tasks_pay_to_inv_error_categories}
\end{figure}

\experiment{Exp. 6: Data errors and task complexity.} 
Our second experiment analyzes the impact of different error types on enterprise tasks like payment-to-invoice matching, highlighting typical data challenges that appear in enterprise scenarios.
We insert individual types of data errors into the clean data from Exp.~5 to study their effects in isolation. We focus on four typical cases: (1)~we deduct up to USD~0.1 from the paid amount, (2)~we remove or change digits in the assignment or (3)~in the billing number, and (4)~we slightly vary the partner name, for example with abbreviations like \emph{KL Technologies} instead of \emph{Kim \& Lee Technologies}.
The results shown in Figure~\ref{fig:tasks_pay_to_inv_error_categories} indicate that even minor discrepancies can lead to performance degradations. 
Interestingly, errors in the numerical fields cause only  modest performance declines, whereas discrepancies in the business partner names result in much more severe performance drops.
This suggests that LLMs rely heavily on the availability of non-erroneous textual fields for entity matching, highlighting their inability to understand the complex numerical and symbolic values prominent in enterprise data.

\subsection{Experiments \& Results (Case Study 2)}
\label{sec:case_study_2}

To study the compound tasks described in C6, we now conduct experiments based on our second case study \emph{(database integration)}. 

\subsubsection*{Setup.}
We use a dataset for the integration of two companies' customer databases that is based on real-world data~(see Appendix~\ref{appendix:case_study_2} in our extended technical report).
Company A's database consists of a single table with 15 columns, selected as a subset of the SAP customer master data table \texttt{KNA1} with columns such as \texttt{KUNNR}, \texttt{NAME1}, and \texttt{ERDAT}. 
Company B's database consists of two tables connected by a foreign key relationship, with descriptive column names such as \texttt{Organization $\!$Name}, \texttt{Address}, and \texttt{Tax $\!$Number}.
We experiment with datasets of different sizes, from 50 to 300 customers, to analyze the effects of the data scale.
The customers in the two databases have an overlap of~60\%, so 40\% of the customers exist only in one of the databases.
To evaluate the predicted result table, we compare it with the ground truth result table of integrated customers and compute the cell-level accuracy (i.e. the fraction of correct cells).

\begin{table}
\caption{Increased task complexity when matching payments to invoices causes a steady decrease in F1 scores. The first column \emph{(Clean)} resembles simple entity matching on data from public sources, while the last column \emph{(+ Multiple Tables)} represents the enterprise scenario with all its complexities.}
\label{tab:tasks_pay_to_inv}
\begin{center}
    \setlength{\tabcolsep}{6pt}
\notesize{
\begin{tabular}{l c *4{S[table-format=1.2]}}
\toprule
\addlinespace[-0.1pt]
&\multicolumn{1}{c}{\textbf{Clean}}&\multicolumn{1}{c}{\textbf{+ Errors}}&\multicolumn{1}{c}{\textbf{+} \thead[l]{Multi-\\Matches}}&\multicolumn{1}{c}{\textbf{+} \thead[l]{Multiple\\Tables}}\\
\addlinespace[-1.8pt]
\midrule
\textbf{GPT-4o-Mini} & 0.98 & 0.58 & 0.53 & 0.45 \\
\textbf{GPT-4o} & 0.97 & 0.80 & 0.64 & 0.58 \\
\textbf{Claude 3.5 Sonnet} & 0.97 & 0.89 & \textbf{0.86} & 0.58 \\
\textbf{Llama 3.1 Instruct} & \textbf{0.99} & \textbf{0.95} & 0.81 & \textbf{0.72} \\
\bottomrule
\end{tabular}
}
\end{center}
\end{table}

\experiment{Exp. 7: Error propagation in compound tasks.} 
\label{exp:compound_error_propagation}
In our first experiment, we analyze the effect of error propagation in compound tasks with multiple steps. 
We formulate the database integration task as a sequence of three steps, as visualized in Figure~\ref{fig:motivation} (center):

\newparagraph
\emph{1. (Schema Matching)}$\:$
To integrate the data, we first perform schema matching to map the columns of Company B's schema to the columns of Company A's schema.
Following existing research, we formulate the task as a binary classification between columns~\cite{narayan_can_2022, parciak24schema_matching}.
Challenges arise because, for example, the address is represented by a single column at Company B, but split into multiple columns at Company A.
We measure the accuracy of this step as the fraction of correctly found matches for the columns in Company A, since the output of the schema matching step is used to transform Company B's database into Company A's schema.

\newparagraph
\emph{2. (Entity Matching)}$\:$
Next, we perform entity matching to determine which customers have entries in both companies' databases. 
Similar to existing literature \cite{narayan_can_2022, peeters_using_2023}, we formulate entity matching as a binary classification task: Given a pair of rows, one from Company A's table and one from Company B's transformed table, the LLM determines whether they refer to the same customer. 

\newparagraph
\emph{3. (Data Integration)}$\:$
To produce the integrated customer table, we use the LLM to merge the duplicate customer records found in both databases into a single row and to transform Company B's remaining customer records into the format of Company A.

\begin{figure}
    \includegraphics[width=\linewidth]{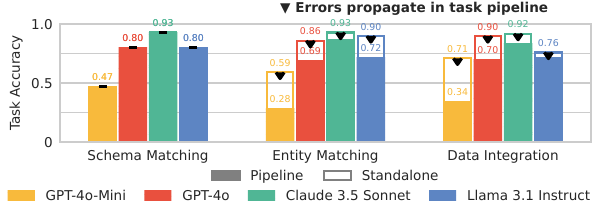}
    \caption{Error propagation in our database integration case study with 100 customers.
    Colored bars represent the \emph{pipeline} setting, where errors carry over from previous steps, whereas stacked colorless bars show the \emph{standalone} setting with correct inputs. The accuracy drop in the pipeline setting highlights the propagation of errors from earlier stages.}
    \label{fig:tasks_compound_pipeline_standalone}
\end{figure}

\newparagraphwithindent
To measure the impact of error propagation when chaining the tasks, we run each step twice: once in a \emph{pipeline} using the output of the previously executed step as input, and once in a \emph{standalone} setting using the ground-truth output of the previous step as input.
Figure~\ref{fig:tasks_compound_pipeline_standalone} shows the results of this experiment.
For the second (entity matching) and third (data integration) step of the pipeline, we observe that executing them in a pipeline leads to a decrease in accuracy compared to running them as standalone tasks. 
However, the decrease depends on the model, with GPT-4o-Mini seeing a more severe decrease than Claude~3.5~Sonnet and Llama~3.1~Instruct.

Interestingly, the accuracy between the entity matching and data integration steps in the pipeline does not decline and even slightly improves for some models.
This shows a surprising effect for LLMs: While the models can make mistakes that propagate through the pipeline, they can, in some cases, also correct earlier mistakes.
For example, some models fail to correctly associate the \texttt{LAND1} field with the address during schema matching.
However, when presented with three sample rows from Company A during data integration, they correctly populate the \texttt{LAND1} field with the country mentioned in the address stored in Company B's data.

Overall, these results suggest that error propagation can lead to a decrease in accuracy for compound tasks, but the significance depends on the LLM.
In some cases, the effects  are negligible. 

\experiment{Exp. 8: End-to-end task execution.}
\label{exp:compound_end2end}
A substantial overhead in enterprise scenarios stems from the fact that complex tasks must be manually decomposed into smaller steps.
Recent LLMs, however, can use reasoning to solve multi-step problems on their own.
One might now ask the question if the manual decomposition of the task is even needed, or if reasoning LLMs can break down the task on their own.
To answer this question, we use the LLMs to directly integrate the two customer databases in a single LLM call and compare the results of this end-to-end execution to the pipelined execution from Exp.~7.
We experiment with two prompting strategies: \textit{w/o steps} provides a textual description with instructions on how to perform the integration but does not mention how to break down the task, whereas \textit{with steps} explicitly mentions the necessary steps (schema matching, entity matching, and data integration) in the prompt to provide a hint towards how to execute them.

\begin{figure}
    \includegraphics[width=\linewidth]{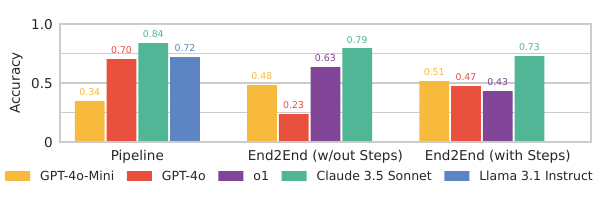}
    \captionsetup{singlelinecheck=false}
    \caption{Pipelined vs. end-to-end execution in our database integration case study with 100 customers. Even at this small scale, end-to-end execution does not outperform pipeline execution.$\!$\protect\footnotemark}
    \label{fig:tasks_compound_pipeline_end2end}
\end{figure}
    \footnotetext{Three bars are missing from this plot: We did not execute the pipeline with o1 due to cost reasons, and Llama~3.1~Instruct did not produce responses for the end-to-end execution because of timeouts.}

\begin{figure}
    \includegraphics[width=\linewidth]{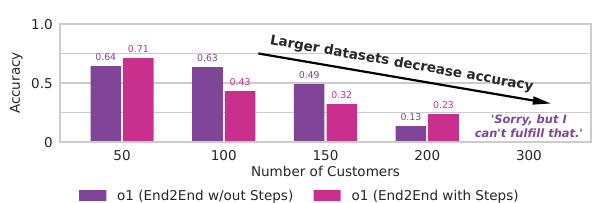}
    \caption{Scaling up the number of customers for end-to-end database integration with OpenAI's o1 drastically decreases the performance. For 300 customers, the model responded \emph{Sorry, but I can't fulfill that.}}
    \label{fig:tasks_compound_end2end_scale}
\end{figure}

Figure~\ref{fig:tasks_compound_pipeline_end2end} shows the results of this experiment, including o1 from OpenAI as an additional LLM with reasoning capabilities.
Most models perform better in the pipelined task execution, with only GPT-4o-Mini performing better in the end-to-end execution.
When comparing the end-to-end execution \textit{w/o steps} and \textit{with steps}, we see mixed results.
Whereas some LLMs benefit from describing the steps in the prompts, others do not.

\experiment{Exp. 9: End-to-end scaling.}
While the end-to-end execution of compound tasks is attractive because it does not require manual task decomposition, it requires the full tables to be provided as input to the LLM because some steps like entity matching might need access to all tuples.
This poses a problem, as larger tables lead to longer prompts that quickly exceed the LLM's context window.
As a last experiment in our database integration case study, we study the effects of scaling the number of customers in the company databases. 
Among the models evaluated in this work, only OpenAI's o1 provides a sufficiently large output token limit (100K tokens) to integrate more than 100 customers in an end-to-end manner.

Figure~\ref{fig:tasks_compound_end2end_scale} shows the results of scaling the total number of customers from 50 up to 300.  
We observe a drastic decline in performance as the number of customers increases, suggesting that the model struggles with long-range dependencies and finding matches in large datasets.
When using 300 customers, the model even responds with \emph{"Sorry, but I can't fulfill that"}. 
Overall, LLMs remain limited in their ability to execute tasks at enterprise-scale in an end-to-end manner.
In contrast, breaking up compound tasks into individual steps can, in principle, scale to larger tables, as tuples are processed one by one (e.g., by comparing all customers of databases A and B individually for entity matching).

\subsection{Discussion}
\label{sec:tasks-discussion}

\subsubsection*{Learnings.}
Through our two case studies, we have demonstrated that automating enterprise tasks with LLMs remains extremely hard.
First, we still need to rely on human effort since end-to-end execution with LLMs is brittle and does not scale well.
This is consistent with \citet{ashury-tahan_mighty_2025}, who found that even strong models often fail to perform robustly on complex table tasks.
We also encountered problems in the output of some task formulations.
For example, during end-to-end execution, where the output is a table with multiple rows, LLMs often generate varying numbers of columns for different rows.

Second, much of the complexity of enterprise tasks is inherent to business needs and cannot be avoided, such as having to deal with multi-match cases in entity matching.
While reducing complexity by working with database views did improve results, constructing such views causes additional manual overheads since business objects in enterprises can be composed of tens or even hundreds of tables.
Finally, some challenges like error propagation are not as severe as initially thought.
Nevertheless, given their current abilities, LLMs are still not sufficient to achieve the level of performance required for enterprise-scale data engineering.

\subsubsection*{Outlook.}
We believe that LLMs out-of-the-box will not be sufficient to address the complexity of enterprise tasks because much of it is inherent to enterprise needs.
Instead, complex tasks call for carefully-designed systems that use LLMs as reasoning or tool-calling agents while effectively incorporating human-LLM interactions.
\citet{wornow_2024_automating_enterprise} propose demonstrating workflows to foundation models, which is promising for automating frequently occurring tasks but still requires manual overheads.
Another promising approach uses LLMs to create structured plans for compound tasks similar to recent advancements in LLMs for query planning~\cite{urban_2024_caesura, liu_palimpzest_2025} before executing these plans step-by-step using existing approaches (e.g., heuristics for schema matching) or even the LLM itself. 
This direction is reinforced by our findings: our experiments show that the stepwise execution of compound tasks improves performance over end-to-end execution while also enabling the LLM to correct earlier mistakes, for example by using information from subsequent steps.

%% file: sections_06_knowledge.tex
\section{The Knowledge Challenge}
\label{sec:knowledge}

Since LLMs are primarily trained on public data, they lack information about a company's internal business processes and policies, as well as about its proprietary tools and systems~\cite{kayali_mind_2025}.
In this section, we analyze how the lack of company-specific knowledge in LLMs affects their accuracy on data engineering tasks.

\subsection{Knowledge Challenges}

We identify two main categories of challenges: the lack of knowledge that is not available to LLMs but does exist in documentation and other company-internal sources, and company-specific extensions of databases that are typically not documented at all.

\challenge{C9: Proprietary but available knowledge.}
Enterprises often use proprietary tools and systems to solve data engineering challenges.
In contrast to well-established technologies like SQL, there is typically much less documentation about them available on the web.
Therefore, it is reasonable to assume that LLMs trained on public data have little parametric knowledge about these tools and systems.
One example is the domain-specific query language SIGNAL~\cite{kampik_signal_2023}, which SAP provides to its customers for exploring data about business processes.
SIGNAL resembles SQL in many aspects but also includes domain-specific features and syntax differences tailored towards process mining.
While the language itself is well-specified on its public documentation page, it has a much smaller user base than SQL and, therefore, a smaller online footprint in terms of help pages, Q\&A threads, and blog posts.
Nevertheless, SIGNAL is frequently used by data engineers and must thus be well-supported by LLMs, for example, to translate natural language requests into SIGNAL queries.

\challenge{C10: Proprietary and unavailable knowledge.}
Enterprise systems like those from SAP support company-specific changes and extensions to customize the system for individual customers.
For example, customers can use hooks to add custom business logic.
On the data level, customization means extending the database schema by adding customer-specific columns to existing tables or even additional tables to the customer namespace.
In our analysis of real-world systems at SAP, we have come across thousands of such customer-defined tables.
Since these changes are highly company-specific, they are typically not documented publicly---if they are documented at all.
Moreover, customers sometimes \enquote{misuse} existing attributes of the standard SAP schema for different purposes.
Such digressions from the public documentation pose significant challenges for data engineering with LLMs as well.

\subsection{Experiments \& Results}

To analyze how challenge C9 affects data engineering with LLMs, we use LLMs to translate natural language requests into the domain-specific query language SIGNAL.
The specifics of SIGNAL provide a good example for the proprietary knowledge primarily available in company-internal documentation.
For challenge C10, we refer to Figure~\ref{fig:headline_exp} as well as an additional experiment in Appendix~\ref{appendix:schema_customization} of our extended technical report.

\experiment{Exp. 10: Text-to-SIGNAL.}
In this experiment, we use LLMs to translate natural language requests into the proprietary query language SIGNAL, as shown in Figure~\ref{fig:motivation} (right).

\subsubsection*{Setup.}
We experiment on a set of 200 randomly-sampled pairs of natural language requests and SIGNAL queries from a larger dataset in use at SAP.
Given a request like \textit{Retrieve the number of unique invoices from 'defaultview-255'}, we prompt the LLM to generate a corresponding SIGNAL query.
We compare three approaches:
First, we provide the LLM only with a short instruction explaining the task \emph{(Zero-shot)}.
Next, we add three fixed example queries that were hand-picked from the SIGNAL documentation \emph{(+ Examples)}.
Finally, we include both the examples as well as the documentation of the SIGNAL language in the prompt \emph{(+ Documentation)}.
We manually tune our prompts in all scenarios to alleviate obvious mistakes and hint at the differences between SIGNAL and SQL.
To evaluate the generated queries, we compute the execution accuracy by executing the ground truth and predicted query and comparing their results.

\subsubsection*{Results.} 
\begin{figure}
    \includegraphics[width=\linewidth]{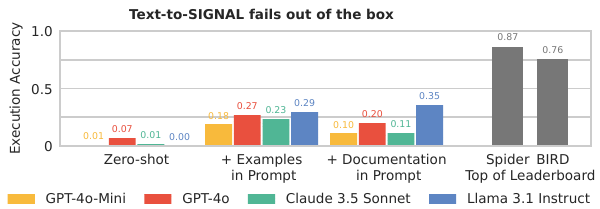}
    \caption{Text-to-SIGNAL execution accuracy. The accuracy remains low even when including hand-picked examples and the SIGNAL documentation in the prompt. By contrast, LLMs achieve high accuracies on Text-to-SQL benchmarks~\cite{gao_text--sql_2024,gao_preview_2025}.}
    \label{fig:knowledge_signal}
\end{figure}

Figure~\ref{fig:knowledge_signal} (left) shows the results of this experiment.
Only a small fraction of the generated SIGNAL queries are correct.
Adding example queries to the prompt brings the biggest improvements in accuracy.
By contrast, additionally including the documentation of the SIGNAL language in the prompt improves the accuracy only  for some models, indicating that simply providing the required documentation during inference is not a viable solution.
Finally, a comparison with the results on popular Text-to-SQL benchmarks~\cite{yu2018spider, li2023birdbench} (Figure~\ref{fig:knowledge_signal} right) shows that the out-of-the-box accuracy for Text-to-SIGNAL is substantially lower.

\begin{figure}
    \includegraphics[width=\linewidth]{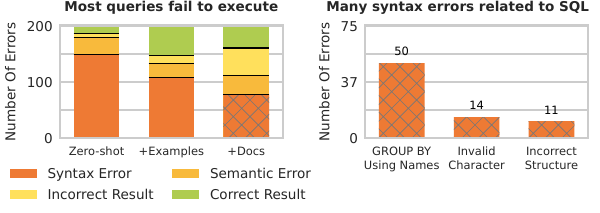}
    \caption{Text-to-SIGNAL errors for GPT-4o. Most queries fail to execute because of syntactic or semantic errors. Examples and documentation in the prompt reduce syntax errors but cause semantic errors, as the model confuses parts of the documentation with the input question. Most syntax errors are caused by small differences between SIGNAL and SQL.}
    \label{fig:knowledge_signal_errors}
\end{figure}

\paragraph{Why are examples and documentation not helping?}
To better understand why exactly the models fail, we conduct an error analysis that categorizes the queries into those that execute with correct or incorrect results, and those that fail to execute because of syntactic or semantic errors.
We focus on GPT-4o, the only model that correctly translated a non-negligible number of requests in all three configurations.
Figure~\ref{fig:knowledge_signal_errors} (left) shows that most of the generated queries fail to execute because of syntactic or semantic errors, with the largest number of failures caused by syntax errors.
Adding example queries and documentation to the prompt causes the number of syntax errors to decrease.
However, the number of queries that contain semantic errors or return incorrect results increases as we add examples and documentation.
A closer inspection of the generated queries reveals that the models often confuse parts of the documentation, like the column names used in explanations, with the request they have to translate.
As a result, the number of fully correct queries is far from satisfactory even when including hand-picked examples and the documentation in the prompt.

\paragraph{A bias towards SQL.}
Another interesting effect is that the models tend to generate SQL-specific syntax and ignore the specifics of SIGNAL.
To further dive into the root causes, we manually analyze the 75 syntax errors generated by GPT-4o in the \emph{(+ Documentation)} setting and group them into three categories: the incorrect use of \emph{GROUP BY} and \emph{ORDER BY}, which in SIGNAL requires numerical indices instead of column names, the use of invalid characters, like an asterisk (*) in a count statement which exists in SQL but not in SIGNAL, and the incorrect structure of the overall statement.
As shown in Figure~\ref{fig:knowledge_signal_errors} (right), we find that even when explicitly including rules in the prompt to avoid such SQL-related syntax errors, the models still generate erroneous SIGNAL queries.
This behavior indicates a strong bias of LLMs towards SQL, which may be caused by its much higher prominence in their training data.
Similar problems could also exist in other scenarios where enterprise data and tasks differ only slightly from public knowledge, like currency conversion or company-specific data types.

\subsection{Discussion}
\label{sec:knowledge-discussion}

\subsubsection*{Learnings.}
Our experiments show that the lack of enterprise-specific knowledge affects the performance of LLMs.
Even when including hand-picked examples and the full documentation in the prompt, LLMs are unable to actualize this information to solve enterprise tasks like Text-to-SIGNAL.
An interesting finding here is that biases in the models' knowledge towards public sources, like their understanding of SQL, are hard to overcome even with explicit instructions, making LLMs difficult to adapt 
without more heavy-weight solutions like fine-tuning that cause additional overheads.

\subsubsection*{Outlook.}
Our results demonstrate that simple out-of-the-box approaches like few-shot prompting and including relevant documentation in the prompt are insufficient.
While we believe that more involved approaches, like fine-tuning LLMs for specific enterprise use cases, could help them acquire the necessary domain knowledge, these approaches require extensive work, for example to label the necessary data for fine-tuning for each individual task.
Moreover, even self-supervised (task-agnostic) pre-training on enterprise corpora comes with important challenges: enterprise data often contains valuable and sensitive information, which raises significant privacy and copyright concerns.
Therefore, we believe that future research should explore controlled alternatives, such as working with synthetic or anonymized corpora that reflect enterprise data characteristics \cite{10184664}.
Despite all these challenges, we believe that LLMs tuned to enterprise settings are a promising direction and ideas in that direction have recently gained traction in industry to adapt foundation models for enterprise environments.\footnote{\url{https://www.databricks.com/company/newsroom/press-releases/databricks-and-anthropic-sign-landmark-deal-bring-claude-models} (last accessed November 4, 2025)}

%% file: sections_07_costs.tex
\section{Costs At Enterprise Scale}
\label{sec:costs}

The cost of using LLMs at enterprise scale is an important orthogonal dimension to the challenges discussed before.
In this section, we want to briefly highlight three main cost drivers: the large scale of the data, the algorithmic complexity of the tasks, and the high tokenizer fertility~\cite{rust_how_2021}, which is a general problem on tabular data.

\paragraph{Data scale.}
As described in Section~\ref{sec:data}, enterprise databases are substantially larger than typical data engineering benchmark datasets, with SAP databases often reaching multiple terabytes in size.
Table~\ref{tab:costs_tokenization} shows how this large size translates into high LLM costs.
Processing one gigabyte of tabular data from our \sapcta corpus with GPT-4o-Mini (OpenAI's cheapest model) already costs about USD~71.
Note that this cost does not include generating any output tokens, which are substantially more expensive than input tokens, as shown in Table~\ref{tab:models}.
Nevertheless, the input token costs already provide a rough estimate of the expected high costs.
For example, processing the entire customer database with GPT-4o-Mini would cost USD~463T, and using the o1 reasoning model would cost USD~46M.
These high costs clearly prohibit using LLMs on large fractions of the enterprise data.
This is especially problematic as our experiments in Sections~\ref{sec:data}-\ref{sec:knowledge} have shown that cheaper models like GPT-4o-Mini often underperform more expensive models.

\paragraph{Algorithmic complexity.}
As a second cost driver, we want to highlight the algorithmic complexity of the ways in which many data engineering tasks are currently approached with LLMs.
One example is the task of entity matching, which current approaches address by using LLMs to compare pairs of entities~\cite{peeters_using_2023,li_table-gpt_2024}.
This approach has an algorithmic complexity of $O\left(N \times M\right)$, where $N$ and $M$ are the numbers of entities in each table, making it intractable even for medium-sized tables.
For example, matching just 1,000 payments to 1,000 invoices already costs USD~1,462, and matching 10,000 payments to 10,000 invoices would cost USD~146,165.
Future work should therefore focus on reducing the cost of LLM-based approaches by rethinking task formulations and combining LLMs with smaller fine-tuned models~\cite{liu_magneto_2024,zhang_directions_2024} and other techniques.
For example, for entity matching, \citet{wang_match_2025} propose a task formulation where, given a row from one table, the LLM is asked to identify all matching rows from the other table in one step.
This reduces the complexity to $O(N)$ if the second table fits in context, and to $O(N \times M/B)$ when batching is used with batch size $B$, offering a substantial reduction in cost without compromising accuracy.

\begin{table}
\caption{Tokens per byte and cost per GB for textual and enterprise data. Enterprise data requires twice as many tokens per byte, leading to twice the cost per GB. Encoding the entire SAP customer database causes high costs for all models.}
\label{tab:costs_tokenization}
\centering
\begin{threeparttable}
\setlength{\tabcolsep}{5.8pt}
  \notesize{
\begin{tabular}{l c rrr}
\toprule
&\multirow{2}{*}{\thead[c]{Tokens\\per Byte}}
&\multicolumn{3}{c}{\textbf{USD per GB}}\\
\cmidrule(lr){3-5}
&&\textbf{GPT-4o-Mini}&\textbf{GPT-4o}&\textbf{o1}\\
\midrule
\textbf{Wikipedia}&0.23&34&574&3,442\\
\textbf{\sapcta}(CSV)&0.47&71&1,181&7,085\\
\midrule
\multicolumn{2}{l}{\textbf{USD for entire database:}}&462,923&7,715,380&46,292,282\\
\bottomrule
\end{tabular}
}
\footnotesize{
\begin{tablenotes}
\hfill Pricing by OpenAI in February 2025.
\end{tablenotes}
}
\end{threeparttable}
\end{table}

\paragraph{Tokenizer fertility.}
A final important cost driver is the high fertility of existing tokenizers on tabular data.
The tokenizer fertility measures the number of tokens required to represent a single piece of text~\cite{rust_how_2021}.
Since tokenizers are trained to represent natural language text, their vocabulary contains tokens even for long words.
By contrast, tabular enterprise data contains lots of symbols and numerical values.
As shown in Table~\ref{tab:costs_tokenization}, this results in twice the number of tokens required to represent each byte of enterprise data compared to natural language text from Wikipedia, leading to twice the cost per gigabyte.
Even if the cost of new models continues to decrease, the costs for processing tabular data will thus remain twice as high.
Therefore, future work should focus on creating new tokenizers specifically designed for enterprise data, as well as on creating more efficient textual representations for such data~\cite{trummer_generating_2024}.

%% file: sections_08_conclusion.tex
\section{Conclusion}
\label{sec:conclusion}

In this paper, we systematically analyzed the challenges involved in applying LLMs to real-world data engineering scenarios in enterprises.
Through experiments with multiple LLMs on a diverse set of tasks, we have shown that the effectiveness of these models is substantially affected by enterprise-specific complexities along the dimensions of \emph{data}, \emph{tasks}, and \emph{knowledge}.
While the absolute numbers of the observed effects may change with more experiments and newer models, the large gap between enterprise and public scenarios shows the clear need for future research.
On the same note, we see a strong need for enterprise-specific LLMs, with many recent activities emerging from research and industry. 
We hope that our insights and learnings provide a helpful guide and inspiration for future efforts to make LLMs viable for enterprise data engineering and support their adoption in industry.

%% file: appendix.tex
\newpage
\appendix

\section{Extended Technical Report}
This extended technical report includes additional information, experiments, and ablation studies. 
All published artifacts for the experiments are provided in our repository.$\!$\footnote{\url{https://github.com/DataManagementLab/llmeval-enterprise-challenges}}

\subsection{Task: Column Type Annotation}
\label{appendix:column_type_annotation}

\paragraph{Table selection.}
When planning our experiments, we consulted experts at SAP to get an overview of the tables in the ERP system and decide which of them are the most important (i.e., used most frequently by customers).
This discussion led us to the selection of 100 tables across 13 SAP business modules covering a broad range of application domains, as shown in Table~\ref{tab:sapcta_construction}.
We chose this process over randomly sampling tables because SAP systems typically contain multiple thousands of tables, many of which are not actively used by each of their customers.
To create the \sapcta dataset, we extracted the expert-selected tables (including their customer-specific modifications) from an actual SAP customer system.

\begin{table}[h]
  \caption{\sapcta dataset breakdown. The dataset contains 100 tables from the 13 most important SAP domains (business modules). The table shows the distribution of the 100 tables over the business modules.}
  \label{tab:sapcta_construction}
  \centering
  \setlength{\tabcolsep}{6pt}
  \notesize{
  \begin{tabular}{l r}
\toprule
\textbf{SAP Business Module}&\textbf{Tables}\\
\midrule
Materials Management (MM)&23\\
Production Planning (PP)&18\\
Sales and Distribution (SD)&15\\
Financial Accounting (FI)&13\\
Quality Management (QM)&6\\
Plant Maintenance (PM)&5\\
Forecasting / MRP&4\\
Warehouse Management (WM)&4\\
Controlling (CO)&4\\
Project System (PS)&3\\
Vendor Master Data (MM-PUR)&2\\
Payment Processing&2\\
Product Grouping&1\\
\bottomrule
\end{tabular}
}
\end{table}

\paragraph{On the number of semantic types in enterprise data.}
There are many more semantic types in \sapcta than in public benchmarks.
The semantic types in \sapcta cover a wide spectrum of business domains, from Financial Accounting and Controlling to Quality Management and Human Resources.
Moreover, SAP needs to cover a broad variety of customer domains, ranging from chemistry and car manufacturing to public administration and education.
For metadata management, SAP provides a data dictionary with descriptive semantic types that we use in our experiments.
These semantic types are often quite fine-granular since they have to describe complex conditions, like \emph{Account Number of Vendor or Creditor} or \emph{Amount Difference in Local Currency}.

By contrast, the semantic types in public benchmarks are often more high-level (e.g., \emph{team}, \emph{definition}, or \emph{cost}).
Many datasets like GitTablesCTA~\cite{madelon_hulsebos_2021_5706316}, SOTAB~~\cite{korini_sotab_2022}, and WikiTables-TURL~\cite{deng_turl_2020} use semantic types from other public resources like DBPedia, Schema.org, and Freebase, and the authors of WikiTables-TURL state ``We further filter out types with less than 100 training instances and keep only the most representative types. In the end, we get a total number of 255 types, [..].'' \cite{deng_turl_2020}.
By deliberately limiting the number of semantic types during dataset creation, the problem is made less difficult compared to if all types would have been used.
On the other hand, the SportsTables dataset with 452 semantic types has more specific semantic types (e.g., \emph{baseball.team.grounded\_double\_plays} or \emph{football.player.rushing\_yards\_per\_game}), but covers only one domain (sports) that is also common knowledge.

To conclude, the \sapcta dataset has a much higher number of semantic types because they must cover many different domains and must be quite specific.
Moreover, they are harder because they are often not common knowledge (e.g., \emph{Product Cost Collector} or \emph{Theil coefficient}) and less likely part of the LLMs' training data than those from web resources.

\paragraph{Ablation Study: Example row selection strategies.}
In our column type annotation experiments, we limit each table to three randomly-selected example rows.
In the following ablation experiments, we investigate the effects of including more example rows (Table~\ref{tab:data_cta_headers_num_rows_appendix}) and experiment with different row selection strategies to reduce the sparsity of the example rows (Table~\ref{tab:data_cta_headers_sampling_modes_appendix}).

\begin{table}
  \caption{Different numbers of example rows. The table shows support-weighted F1 scores for column type annotation with and without column names. Including more example rows in the prompt does not improve results on enterprise data.}
  \label{tab:data_cta_headers_num_rows_appendix}
  \centering
  \setlength{\tabcolsep}{4pt}
  \footnotesize{
  \begin{tabular}{l *6{S[table-format=1.2]}}
\toprule
&\multicolumn{2}{c}{\textbf{GitTablesCTA}}
&\multicolumn{2}{c}{\textbf{SportsTables}}
&\multicolumn{2}{c}{\textbf{\sapcta}}
\\
\cmidrule(lr){2-3} \cmidrule(lr){4-5} \cmidrule(lr){6-7}
\textbf{Column Names}&\textbf{w/out}&\textbf{with}&\textbf{w/out}&\textbf{with}&\textbf{w/out}&\textbf{with}\\
\midrule
\textbf{GPT-4o-Mini} \\
\textbf{\quad \underline{3 rows}} & 0.52 & \textbf{0.96} & \textbf{0.27} & \textbf{0.55} & \textbf{0.02} & \textbf{0.07} \\
\textbf{\quad 5 rows} & \textbf{0.53} & \textbf{0.96} & \textbf{0.27} & 0.51 & \textbf{0.02} & \textbf{0.07} \\
\textbf{\quad 10 rows} & \textbf{0.53} & \textbf{0.96} & \textbf{0.27} & 0.48 & 0.01 & \textbf{0.07} \\
\textbf{\quad 20 rows} & 0.52 & \textbf{0.96} & 0.25 & 0.42 & 0.01 & 0.06 \\
\midrule
\textbf{GPT-4o} \\
\textbf{\quad \underline{3 rows}} & \textbf{0.56} & \textbf{0.99} & 0.57 & \textbf{0.91} & 0.03 & 0.24 \\
\textbf{\quad 5 rows} & 0.52 & \textbf{0.99} & 0.59 & \textbf{0.91} & \textbf{0.04} & 0.23 \\
\textbf{\quad 10 rows} & 0.52 & \textbf{0.99} & 0.62 & \textbf{0.91} & \textbf{0.04} & 0.23 \\
\textbf{\quad 20 rows} & 0.52 & 0.98 & \textbf{0.64} & 0.89 & 0.03 & \textbf{0.26} \\
\bottomrule
\end{tabular}
}
\end{table}

\begin{table}
  \caption{Different example row selection strategies. The table shows support-weighted F1 scores for column type annotation with and without column names. The example row selection strategies do not improve results on enterprise data.}
  \label{tab:data_cta_headers_sampling_modes_appendix}
  \centering
  \setlength{\tabcolsep}{2.8pt}
  \footnotesize{
  \begin{tabular}{l *6{S[table-format=1.2]}}
\toprule
&\multicolumn{2}{c}{\textbf{GitTablesCTA}}
&\multicolumn{2}{c}{\textbf{SportsTables}}
&\multicolumn{2}{c}{\textbf{\sapcta}}
\\
\cmidrule(lr){2-3} \cmidrule(lr){4-5} \cmidrule(lr){6-7}
\textbf{Column Names}&\textbf{w/out}&\textbf{with}&\textbf{w/out}&\textbf{with}&\textbf{w/out}&\textbf{with}\\
\midrule
\textbf{GPT-4o-Mini} \\
\textbf{\quad \underline{3 random rows}} & 0.52 & 0.96 & \textbf{0.27} & \textbf{0.55} & 0.01 & \textbf{0.07} \\
\textbf{\quad 3 rows with lowest sparsity} & \textbf{0.54} & 0.96 & 0.26 & 0.53 & \textbf{0.02}& \textbf{0.07} \\
\textbf{\quad 3 rows of non-sparse values} & \textbf{0.54} & \textbf{0.97} & 0.25 & 0.53 & 0.01 & \textbf{0.07} \\
\midrule
\textbf{GPT-4o} \\
\textbf{\quad \underline{3 random rows}} & 0.56 & \textbf{0.99} & 0.57 & \textbf{0.91} & \textbf{0.04} & \textbf{0.26} \\
\textbf{\quad 3 rows with lowest sparsity} & 0.55 & \textbf{0.99} & \textbf{0.60} & \textbf{0.91} & \textbf{0.04} & 0.23 \\
\textbf{\quad 3 rows of non-sparse values} & \textbf{0.57} & \textbf{0.99} & 0.54 & 0.88 & 0.03 & 0.24 \\
\bottomrule
\end{tabular}
}
\end{table}

We find that neither including more rows, nor selecting the rows differently improves the results.
We explain this as follows:
When including the column names in the prompt (i.e., \emph{with}), the LLM focuses almost exclusively on them, an observation we already made earlier.
When not including the column names (i.e., \emph{w/out}), we see only slight gains on the SportsTables dataset, but not on the other datasets.
A potential explanation could be that there is a large gap between columns that are easy to label even based on only three values (e.g., company or material names), and those that are hard to label even with a larger number of examples.
For example, integer foreign keys are almost impossible to label correctly even with hundreds of example values, so providing more data does not provide a better signal.
Therefore, for the main paper, we decided to report results only for three example rows.

\paragraph{Ablation Study: Variability of results.}
In the main paper, we report results based on one run of each experiment.
In this ablation study, we investigate the variability of results by conducting three identical runs of the same experiment (Exp.~1).
As shown in Table~\ref{tab:data_cta_headers_mulitple_runs}, the standard deviation between the different runs is low compared to our effect sizes.
Furthermore, Table~\ref{tab:tasks_pay_to_inv_standard_errors} shows bootstrapped standard errors (based on Monte Carlo Case Resampling) for our payment-to-invoice matching case study. The standard errors are all low, indicating that the sample size used for the experiment is large enough to be meaningful.

\begin{table}
  \caption{Variability of results. The table shows support-weighted F1 scores and their standard deviations for column type annotation with and without column names. The standard deviations are low compared to the effect sizes}
  \label{tab:data_cta_headers_mulitple_runs}
  \centering
  \setlength{\tabcolsep}{3.5pt}
  \footnotesize{
  \begin{tabular}{ll *6{S[table-format=1.4]}}
\toprule
&&\multicolumn{2}{c}{\textbf{GitTablesCTA}}
&\multicolumn{2}{c}{\textbf{SportsTables}}
&\multicolumn{2}{c}{\textbf{\sapcta}}
\\
\cmidrule(lr){3-4} \cmidrule(lr){5-6} \cmidrule(lr){7-8}
\multicolumn{2}{l}{\textbf{Column Names}}&\textbf{w/out}&\textbf{with}&\textbf{w/out}&\textbf{with}&\textbf{w/out}&\textbf{with}\\
\midrule
\multicolumn{2}{l}{\textbf{GPT-4o-Mini (Std. dev. ±)}} & 0.0083 & 0.0013 & 0.0015 & 0.0063 & 0.0004 & 0.0009 \\
\cmidrule(){2-8}
&\textbf{Run 1} & 0.5267 & 0.9741 & 0.2886 & 0.5303 & 0.0146 & 0.0710 \\
&\textbf{Run 2} & 0.5431 & 0.9716 & 0.2902 & 0.5179 & 0.0147 & 0.0693 \\
&\textbf{Run 3} & 0.5369 & 0.9731 & 0.2873 & 0.5261 & 0.0153 & 0.0702 \\
\midrule
\multicolumn{2}{l}{\textbf{GPT-4o (Std. dev. ±)}} & 0.0014 & 0.0003 & 0.0057 & 0.0033 & 0.0022 & 0.0162 \\
\cmidrule(){2-8}
&\textbf{Run 1} & 0.5611 & 0.9874 & 0.5845 & 0.9057 & 0.0406 & 0.2427 \\
&\textbf{Run 2} & 0.5627 & 0.9868 & 0.5847 & 0.9073 & 0.0372 & 0.2715 \\
&\textbf{Run 3} & 0.5600 & 0.9868 & 0.5747 & 0.9009 & 0.0364 & 0.2442 \\
\bottomrule
\end{tabular}
}
\end{table}

\paragraph{Public CTA dataset with enterprise challenges.}
The \sapcta dataset used in our experiments consists of tables extracted from an SAP customer database, and we are legally not allowed to publish this customer data. 
To enable others to reproduce our findings and, more importantly, contribute to further advancing LLMs for data engineering in enterprises, we have created a dataset that has the same characteristics as the enterprise data in \sapcta$\!$.
Moreover, we demonstrate that the data we publish leads to similar performance decreases if used by LLMs.

We publish an adapted version of the SportsTables dataset~\cite{langenecker_sportstables_2023} consisting of 100 tables annotated with semantic types in our repository.
In the following, we first describe how we ``infused'' the enterprise data challenges into the dataset before discussing the column type annotation results on this modified dataset.

\begin{figure*}
    \includegraphics[width=0.95\linewidth]{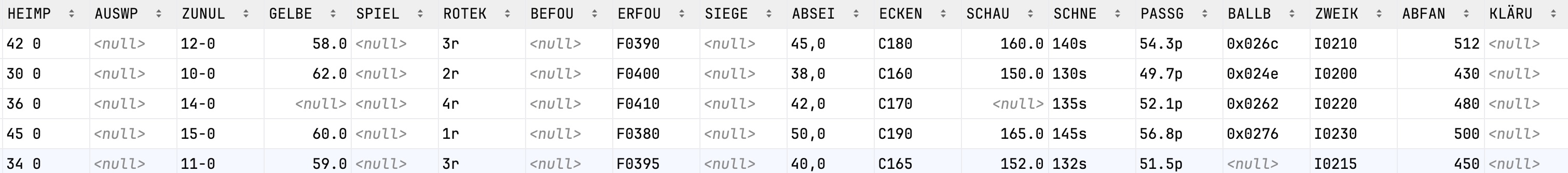}
    \caption{Excerpt of an adapted SportsTables table with all enterprise data challenges (C1-C4 from Section \ref{sec:data_challenges}) applied. The example table has 108 columns, not all of which are shown here.}
    \label{fig:adapted_sportstables}
\end{figure*}

\newparagraph
We apply one dataset transformation for each data challenge:
\begin{enumerate}
    \item \textbf{\textit{Increase table widths:}} For each SportsTables table, we first sample a target number of columns from a beta distribution resembling the distribution of table widths in \sapcta$\!$. To reach the target number of columns, we prompt GPT-4o to generate a list of unique column names to augment the original table. In a second step, we use GPT-4o again to generate the data for these additional columns in chunks of 20 columns. We include the original table in each prompt and instruct the LLM to generate qualitatively similar data to ensure consistency across the table columns.
    \item \textbf{\textit{Lower the descriptiveness of the schemas:}} Next, we lower the descriptiveness of the table and column names to resemble those in \mbox{\sapcta$\!$.} Since the table and column names of SAP tables are typically 4- and 5-letter abbreviations of German concepts, we first prompt GPT-4o to translate the table and column names of the SportsTables dataset into German, again using chunks of 20 columns. In a second step, we prompt GPT-4o again to generate 4- and 5-letter abbreviations of the translated table and column names.
    \item \textbf{\textit{Increase the data complexity (use complex data types):}} To adapt the data values of the SportsTables dataset, we use GPT-4o to transform the data in chunks of 20 columns based on observations made about the data in \sapcta$\!$. For example, we instruct the model to replace explicit textual values like names with numerical and alpha-numerical ids, or to transform numerical values by prepending leading zeros. To ensure consistency in each column as well as a balanced mixture of transformations across columns, the model first plans which transformation makes sense for which column before transforming the actual data.
    \item \textbf{\textit{Increase sparsity of data:}} Finally, we increase the sparsity to match that of \sapcta by masking out cell values.
\end{enumerate}
We create five versions of the transformed dataset: one for each of the challenges C1-C4 and one with all of them applied.
Figure~\ref{fig:adapted_sportstables} shows an excerpt of an adapted table with all challenges applied.

\newparagraph
Table~\ref{tab:data_cta_headers_pub} shows the results of replicating Table~\ref{tab:data_cta_headers} on this adapted version of the SportsTables dataset.
We can clearly see the stark performance decrease caused by the applied enterprise data characteristics.
Moreover, we can see how each data characteristic contributes to the overall performance decrease, indicating that our modifications to the SportsTables dataset indeed reflect the primary challenges of enterprise data.
As shown in Table~\ref{tab:costs_tokenization_imdb}, the dataset also reflects the cost characteristics of enterprise data.

\begin{table}
  \caption{Adapted SportsTables dataset vs. \sapcta$\!$. The table shows support-weighted F1 scores for column type annotation with and without column names. The accuracy on both datasets is highly similar, indicating that both datasets pose similar challenges to LLMs.}
  \label{tab:data_cta_headers_pub}
  \centering
  \setlength{\tabcolsep}{5pt}
  \footnotesize{
  \begin{tabular}{ll *4{S[table-format=1.2]}}
\toprule
&&\multicolumn{2}{c}{\textbf{GPT-4o-Mini}}&\multicolumn{2}{c}{\textbf{GPT-4o}}\\
\cmidrule(lr){3-4} \cmidrule(lr){5-6}
\multicolumn{2}{l}{\textbf{Column Names}}&\textbf{w/out}&\textbf{with}&\textbf{w/out}&\textbf{with}\\
\midrule
\multicolumn{2}{l}{\textbf{SportsTables (Original)}} & 0.27 & 0.55 & 0.57 & 0.91 \\
\cmidrule(){2-6}
&\textbf{With large table widths} & 0.15 & 0.31 & 0.25 & 0.60 \\
&\textbf{With high sparsity} & 0.07 & 0.50 & 0.21 & 0.86 \\
&\textbf{With low descriptiveness} & 0.29 & 0.25 & 0.62 & 0.74 \\
&\textbf{With complex data types} & 0.06 & 0.50 & 0.17 & 0.90 \\
\cmidrule(){2-6}
&\textbf{With all above challenges} & 0.02 & 0.08 & 0.04 & 0.42 \\
\midrule
\multicolumn{2}{l}{\textbf{\sapcta}} & 0.02 & 0.07 & 0.04 & 0.24 \\
\bottomrule
\end{tabular}
}
\end{table}

\begin{table}
\caption{Cost characteristics replicated on the adapted SportsTables dataset, which shows characteristics similar to \sapcta (higher tokenizer fertility, cost per GB, etc.).}
\label{tab:costs_tokenization_imdb}
\centering
\begin{threeparttable}
\setlength{\tabcolsep}{3pt}
  \footnotesize{
\begin{tabular}{l c rrr}
\toprule
&\multirow{2}{*}{\thead[c]{Tokens\\per Byte}}
&\multicolumn{3}{c}{\textbf{USD per GB}}\\
\cmidrule(lr){3-5}
&&\textbf{GPT-4o-Mini}&\textbf{GPT-4o}&\textbf{o1}\\
\midrule
\textbf{Wikipedia}&0.23&34&574&3,442\\
\textbf{\sapcta}(CSV)&0.47&71&1,181&7,085\\
\textbf{SportsTables with C1-C4} (CSV)&0.56&84&1,392&8,350\\
\bottomrule
\end{tabular}
}
\footnotesize{
\begin{tablenotes}
\hfill Pricing by OpenAI in February 2025.
\end{tablenotes}
}
\end{threeparttable}
\end{table}

\subsection{Task: Entity Matching (Case Study 1)}
\label{appendix:case_study_1}

\paragraph{Dataset details.}
We publish our payment-to-invoice matching dataset in our repository.
It consists of four database tables: \texttt{PKPF}, \texttt{BSEG}, and \texttt{KNA1} for invoices, and \texttt{FEBEP} for payments.
The dataset also includes a ground truth table with information about the true matches, including which errors (e.g., incorrect reference number) and which multi-match cases (1:1, 1:N, or N:1) apply to each of them.
The dataset has multiple versions: with up to one or up to multiple errors per match, and with descriptive, non-descriptive, or multi-table schemas.

\begin{table}
\caption{Bootstrapped standard errors (based on Monte Carlo Case Resampling). The table shows F1 scores when matching payments to invoices. The standard errors are low, indicating that our sample size is large enough.}
\label{tab:tasks_pay_to_inv_standard_errors}
\begin{center}
    \setlength{\tabcolsep}{3.5pt}
\footnotesize{
\begin{tabular}{l cccc}
\toprule
\addlinespace[-0.1pt]
&\multicolumn{1}{c}{\textbf{Clean}}&\multicolumn{1}{c}{\textbf{+ Errors}}&\multicolumn{1}{c}{\textbf{+} \thead[l]{Multi-\\Matches}}&\multicolumn{1}{c}{\textbf{+} \thead[l]{Multiple\\Tables}}\\
\addlinespace[-1.8pt]
\midrule
\textbf{GPT-4o-Mini} & 0.98 ± 0.01 & 0.58 ± 0.04 & 0.53 ± 0.03 & 0.45 ± 0.02 \\
\textbf{GPT-4o} & 0.97 ± 0.01 & 0.80 ± 0.03 & 0.64 ± 0.02 & 0.58 ± 0.02 \\
\textbf{Claude 3.5 Sonnet} & 0.97 ± 0.02 & 0.94 ± 0.01 & 0.91 ± 0.01 & 0.64 ± 0.02 \\
\textbf{Llama 3.1 Instruct} & 0.99 ± 0.01 & 0.95 ± 0.01 & 0.81 ± 0.02 & 0.72 ± 0.02 \\
\bottomrule
\end{tabular}
}
\end{center}
\end{table}

\paragraph{Creating views for entity matching.}
As described in Section~\ref{sec:task_challenges}, entities in enterprise systems are often represented by multiple rows stored in different connected tables.
To evaluate how this affects data engineering with LLMs, in our payment-to-invoice matching case study, we compare representing entities as a manually-curated \emph{flat} view to representing them as \emph{multiple separate tables}.
In the following, we describe both settings in more detail.

In the flat setup used in the \emph{(Clean)}, \emph{(+ Errors)}, and \emph{(+ Multi-Matches)} settings of the study, the LLM receives rows of one flat table of invoices and one table of payments.
The flat view of the invoices is constructed by joining three tables from the underlying ERP system where invoice data is stored: \texttt{BKPF} (accounting document header), \texttt{BSEG} (accounting document segment), and \texttt{KNA1} (customer master table).
For the payments, there is only a single table in our dataset which stores payment information, so we directly use it as the flat view of the payments.
To summarize, in the \emph{flat} setting, each LLM prompt contains one row from the payment table and one row from the invoice view described above.

In the \emph{(+ Multiple Tables)} setting of the study, we directly use the base tables as input to the LLM.
Each prompt to the LLM includes one payment row to represent the payment, and one row from \texttt{BKPF}, the corresponding row from \texttt{BSEG}, and the full \texttt{KNA1} table separately for representing the invoice.
Therefore, the LLM needs to operate directly on the multi-table database schema instead of the manually-curated flat view.

\begin{figure}
    \includegraphics[width=\linewidth]{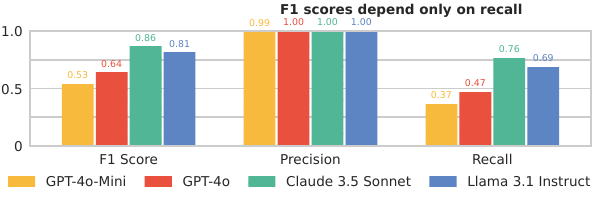}
    \caption{Precision and recall for the \textit{(\textbf{+ Multi-Matches})} setting of our payments-to-invoice matching case study.
    All models achieve very high precision. However, their low recall leads to many false negatives, increasing the manual effort in real-world applications.}
    \label{fig:tasks_pay_to_inv_precision_recall}
\end{figure}

\paragraph{Drill-down into precision and recall.}
Figure \ref{fig:tasks_pay_to_inv_precision_recall} shows the precision and recall for the \emph{(+ Multi-Matches}) setting of our payment-to-invoice matching case study.
All models achieve very high precision.
However, their low recall leads to many false negatives, which increase the manual effort in real-world applications.

\subsection{Task: Database Integration (Case Study 2)}
\label{appendix:case_study_2}

\paragraph{Dataset details.}
We publish our enterprise database integration dataset in our repository.
It consists of the input tables from Company A and Company B, the ground truth output table of the database integration, and the ground truths of the intermediate steps (schema matching and entity matching).
The dataset has multiple versions with sizes between 50 and 300 customers.

\paragraph{LLM self-correction in compound tasks.}
In the pipelined execution of our database integration case study in Section~\ref{exp:compound_error_propagation}, we have seen that LLMs can sometimes correct mistakes made in earlier pipeline steps later on. 
In the following, we provide an example where the LLM fixes a mistake in the third step (data integration) that it made in the first step (schema matching).

\newparagraph
\emph{Schema matching errors:}
As shown in Table~\ref{tab:compound_task_error_sm}, the LLM failed to detect the match between the \texttt{Address1} column of Company~B and the \texttt{LAND1} column of Company~A (\texttt{Address1} contains the information for \texttt{LAND1}, \texttt{STRAS}, \texttt{ORT01}, and \texttt{PSTLZ}).
It also failed to detect the match between the \texttt{TAX Number} column of Company~B and the \texttt{STCEG} column of Company~A.
Because of these errors, the rows of Company~B get transformed incorrectly into the schema of Company~A, as shown in Table~\ref{tab:compound_task_error_incorrect_rows}.

\newparagraph
\emph{Fixed errors during data integration:}
In the data integration step, the model receives example rows (see Table~\ref{tab:compound_task_error_example_rows}) of Company~A in the prompt and is instructed to transform the data in the transformed row of Company~B accordingly.
As shown in Table~\ref{tab:compound_task_error_fixed_errors}, the example rows lead the model to correctly move the country information into the \texttt{LAND1} column (and derive the country code), as well as to correctly determine whether the \texttt{TAX Number} belongs into the \texttt{STCD1} or \texttt{STCEG} column, which must be decided on a case-by-case basis based on the location of the company.

\begin{table*}
\centering
\caption{Example for LLM self-correction in our database integration case study.}
\subcaptionbox{During the first step (schema matching), the LLM fails to detect two matches.\label{tab:compound_task_error_sm}}{
 \setlength{\tabcolsep}{15pt}
  \footnotesize{
  \begin{tabular}{l l}
\toprule
\textbf{Columns of Company B}&\textbf{Mapped to Columns of Company A}\\
\midrule
Address1 & STRAS, ORT01, PSTLZ, \textcolor{red}{LAND1 (missing)} \\
Organization Name &NAME1 \\
Country Prefix, Contact Number & TELF1 \\
Email Address & SMTP\_ADDR \\
Creation Date & ERDAT \\
Modification Date & UPDAT \\
TAX Number & STCD1, \textcolor{red}{STCEG (missing)} \\
\bottomrule
\end{tabular}
}
}
\medskip
\vspace{2ex}
\subcaptionbox{Because of the schema matching errors, the rows of Company~B get transformed incorrectly into the schema of Company~A.\label{tab:compound_task_error_incorrect_rows}}{
 \setlength{\tabcolsep}{4.5pt}
  \tiny{
  \begin{tabular}{lp{1.2cm}p{1.2cm}p{1.2cm}p{1.2cm}p{1.2cm}p{1.2cm}p{1.2cm}p{1.2cm}p{1.2cm}p{1.2cm}p{1.2cm}r}
\toprule
\textbf{...} & \textbf{LAND1} & \textbf{NAME1} & \textbf{STRAS} & \textbf{ORT01} & \textbf{PSTLZ} & \textbf{TELF1} & \textbf{SMTP\_ADDR} & \textbf{ERDAT} & \textbf{UPDAT} & \textbf{STCEG} & \textbf{STCD1} & \textbf{...} \\
\midrule
& \textcolor{red}{(missing)} & Oleo Puro & Acceso Andrés Felipe Amaya 83 07652 Teruel, Spain & Acceso Andrés Felipe Amaya 83 07652 Teruel, Spain & Acceso Andrés Felipe Amaya 83 07652 Teruel, Spain & 34 8205521 & contacto@ oleopuro.com & 2001-04-19 & 2022-12-08 & \textcolor{red}{(missing)} & \textcolor{red}{ESO02430534 (value in wrong column)} & \\
\bottomrule
\end{tabular}
}
}
\medskip
\vspace{2ex}
\subcaptionbox{In the third step (data integration), the model receives example rows of Company~A in the prompt.\label{tab:compound_task_error_example_rows}}{
 \setlength{\tabcolsep}{4.5pt}
  \tiny{
  \begin{tabular}{lp{1.2cm}p{1.2cm}p{1.2cm}p{1.2cm}p{1.2cm}p{1.2cm}p{1.2cm}p{1.2cm}p{1.2cm}p{1.2cm}p{1.2cm}r}
\toprule
\textbf{...} & \textbf{LAND1} & \textbf{NAME1} & \textbf{STRAS} & \textbf{ORT01} & \textbf{PSTLZ} & \textbf{TELF1} & \textbf{SMTP\_ADDR} & \textbf{ERDAT} & \textbf{UPDAT} & \textbf{STCEG} & \textbf{STCD1} & \textbf{...} \\
\midrule
& NL & NovaTerra Real Estate & Mauritsstraat 23 & Dalerpeel & 4019IF & +31 39543992 & info@ novaterra.nl & 20041230 & 20160129 & NL015201365B89 & & \\
& BR & VivaSabor Foods & Núcleo de da Conceição, 79 & Garcia & 50739-219 & +55 62370460 & info@ vivasabor.com.br & 20030914 & 20110911 & & F65216695691895 & \\
\bottomrule
\end{tabular}
}
}
\medskip
\vspace{2ex}
\subcaptionbox{During the third step (data integration), the example rows (see Table~\ref{tab:compound_task_error_example_rows}) lead the model to correct the mistakes it made in the first step (schema matching) by moving the country information and tax number into the correct columns.\label{tab:compound_task_error_fixed_errors}}
{
 \setlength{\tabcolsep}{4.5pt}
  \tiny{
  \begin{tabular}{lp{1.2cm}p{1.2cm}p{1.2cm}p{1.2cm}p{1.2cm}p{1.2cm}p{1.2cm}p{1.2cm}p{1.2cm}p{1.2cm}p{1.2cm}r}
\toprule
\textbf{...} & \textbf{LAND1} & \textbf{NAME1} & \textbf{STRAS} & \textbf{ORT01} & \textbf{PSTLZ} & \textbf{TELF1} & \textbf{SMTP\_ADDR} & \textbf{ERDAT} & \textbf{UPDAT} & \textbf{STCEG} & \textbf{STCD1} & \textbf{...} \\
\midrule
& \textcolor{teal}{ES} & Oleo Puro & Acceso Andrés Felipe Amaya 83 & Teruel & 7652 & +34 8205521 & contacto@ oleopuro.com & 20010419 & 20221208 & \textcolor{teal}{ESO02430534} & & \\
\bottomrule
\end{tabular}
}
}
\end{table*}

\subsection{Task: Text-to-SIGNAL}
\label{appendix:text2signal}

\paragraph{Dataset details.}
We publish our Text-to-SIGNAL dataset in our repository.
It consists of 200 natural language-SIGNAL pairs, each of which consists of a natural language user request and the corresponding ground truth SIGNAL query.
We also publish the prompts, examples, and documentation used in our experiments as well as the generated SIGNAL queries with the annotated error types reported in our experiments.

\balance

\subsection{Proprietary Knowledge: Experiment on Schema Customizations}
\label{appendix:schema_customization}

In this additional experiment, we further analyze the limitations of LLMs regarding company-specific knowledge for which there is little to no documentation, as described in challenge C10.
To demonstrate these limitations, we probe the LLMs' parametric knowledge about the schema of an SAP customer database.
Since LLMs have likely seen the publicly-available SAP documentation during their training, they should have some knowledge about the standard SAP schema, such as which columns belong to which table.
In contrast, they should have no information about customer-specific modifications to the standard SAP schema since these are not documented publicly, and thus information about such columns is highly unlikely to be in their training data.
With this experiment, we want to demonstrate that these customer-defined columns are even harder to understand for LLMs than the standard enterprise data.

\subsubsection*{Setup.}
To mitigate any task-specific difficulties, we choose a simple probing setup:
Given the name of a table in the SAP customer database, we instruct the model to generate a list of all columns that belong to that table.
In our analysis, we differentiate between columns that are part of the standard SAP schema and those that were defined by the customer.
Our corpus consists of 2,000 tables from a real-world SAP system, of which 1,000 are part of the standard SAP schema and 1,000 were defined by the customer.
We evaluate the predictions by computing F1 scores that measure the overlap between the predicted and true columns of each table.

\subsubsection*{Results.} 
\begin{figure}
    \includegraphics[width=\linewidth]{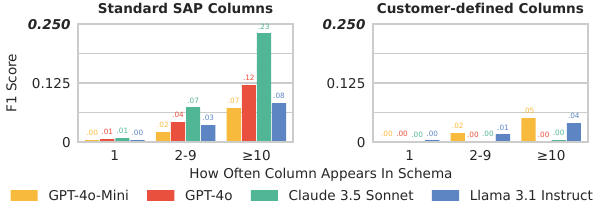}
    \caption{Schema customizations are not understood by LLMs. The plots show F1 scores when predicting columns of standard SAP tables (left) and customer-defined tables (right). The models recall some prominent columns of the standard SAP schema, but do not recall customer-defined columns.}
    \label{fig:knowledge_headers}
\end{figure}

Figure~\ref{fig:knowledge_headers} shows the results broken down by the frequency of how often columns appear in the schema.
As expected, the parametric knowledge about customer-defined columns is severely limited to certain columns like \texttt{MANDT} that must appear in most customer-defined tables.
In contrast, the LLMs achieve higher F1 scores for columns from standard SAP tables.
Still, we see that their parametric knowledge is strongly biased towards columns that appear in many tables and are thus more prominent in public documentation.
Importantly, our introductory experiment (Figure~\ref{fig:headline_exp} right) shows that this lack of company-specific knowledge about customer-defined columns severely impacts performance on downstream tasks like column type annotation.

